\documentclass[12pt,twoside]{article}
\usepackage[mathscr]{eucal}
\usepackage{amsmath,amsfonts,amssymb,amsthm,mathabx}
\usepackage{times}
\usepackage{pdfsync}
\usepackage{cite}
\usepackage{url}
\usepackage{hyperref}

\advance\textheight\topskip
\newcommand{\dd}{\partial}
\renewcommand{\d}{\partial}

\newcommand{\half}{\frac{1}{2}}

\newcommand{\ffrac}[2]{\raisebox{.5pt}%
  {\footnotesize$\displaystyle\frac{#1}{#2}$}\kern1pt}

\newcommand{\ddl}[2]{\ffrac{\dd #1}{\dd #2}}

\def\cH{\mathcal{H}}

\def\cL{\mathcal{L}}

\def\cU{\mathcal{U}}

\numberwithin{equation}{section} \makeatletter
\@addtoreset{equation}{section}

\DeclareFontFamily{OT1}{rsfs}{} \DeclareFontShape{OT1}{rsfs}{m}{n}{
<-7> rsfs5 <7-10> rsfs7 <10-> rsfs10}{}
\DeclareMathAlphabet{\mycal}{OT1}{rsfs}{m}{n}
\def\scri{{\mycal I}}%
\def\scrip{\scri^{+}}%

\newcommand*\xbar[1]{%
  \hbox{%
    \vbox{%
      \hrule height 0.5pt % The actual bar
      \kern0.3ex%         % Distance between bar and symbol
      \hbox{%
        \kern-0.0em%      % Shortening on the left side
        \ensuremath{#1}%
        \kern-0.0em%      % Shortening on the right side
      }%
    }%
  }%
}

\begin{document}

\title{Finite BMS transformations}

\author{Glenn Barnich and C\'edric Troessaert}

\date{}

\def\mytitle{Finite BMS transformations}

\pagestyle{myheadings} \markboth{\textsc{\small G.~Barnich,
    C. Troessaert}}{%
   \textsc{\small Finite BMS transformations}}

\addtolength{\headsep}{4pt}

\begin{centering}

  \vspace{1cm}

  \textbf{\Large{\mytitle}}

  \vspace{1.5cm}

  {\large Glenn Barnich$^a$ and C\'edric Troessaert$^b$}

\vspace{.5cm}

\begin{minipage}{.9\textwidth}\small \it  \begin{center}
   $^a$Physique Th\'eorique et Math\'ematique \\ Universit\'e Libre de
   Bruxelles and International Solvay Institutes \\ Campus
   Plaine C.P. 231, B-1050 Bruxelles, Belgium
 \end{center}
\end{minipage}

\vspace{.5cm}

\begin{minipage}{.9\textwidth}\small \it  \begin{center}
    $^b$Centro de Estudios Cient\'{\i}ficos (CECs) \\
    Arturo Prat 514, Valdivia, Chile
 \end{center}
\end{minipage}

\end{centering}

\vspace{1cm}

\begin{center}
  \begin{minipage}{.9\textwidth}
    \textsc{Abstract}. The action of finite BMS and Weyl
    transformations on the gravitational data at null infinity is
    worked out in three and four dimensions in the case of an
    arbitrary conformal factor for the boundary metric induced on
    Scri.
  \end{minipage}
\end{center}

\vfill

\thispagestyle{empty}
\newpage

\begin{small}
{\addtolength{\parskip}{-2pt}
 \tableofcontents}
\end{small}
\thispagestyle{empty}
\newpage

\section{Introduction}
\label{sec:introduction}

There are two main applications of two dimensional conformal
invariance \cite{Belavin:1984vu}. The first consists in using Ward
identities associated to infinitesimal symmetry transformations in
order to constrain correlation functions. In the second application,
starting from known quantities in a given domain, the finite
transformations are used to generate the corresponding quantities
pertaining to the transformed domain (see
e.g.~\cite{Itzykson:1989sy}). In this case, the Schwarzian derivative
occuring in the transformation law of the energy-momentum tensor plays
a crucial role.

For four-dimensional asymptotically flat spacetimes at null infinity,
an extension of the globally well-defined symmetry group
\cite{Bondi:1962px,Sachs1962a,Sachs1962} in terms of locally defined
infinitesimal transformations has been proposed and studied in
\cite{Barnich:2009se,Barnich:2010eb,Barnich:2011ct,Barnich:2011mi,%
Barnich:2013axa}. In particular, their relevance for gravitational
scattering has been conjectured. Physical implications in terms of
Ward identities for soft gravitons have subsequently been developed in
\cite{Strominger:2013jfa,Cachazo:2014fwa,He:2014laa,Kapec:2014opa}.

The aim of the present paper is to derive the finite transformations
necessary for the second application. In particular for instance, if
one knows the theory in the form of an asymptotic solution to
classical general relativity for the standard topology $S^2\times
\mathbb R$ of $\scri^+$, one can use the transformation laws to get
the solution on a cylinder times a line. Particular aspects of such
mappings in general relativity have been discussed previously for
instance in \cite{Penrose1967,Foster1978,Foster1987}. More concretely,
in the present paper we will work out the transformation laws of
asymptotic solution space and the analog of the Schwarzian derivative
for finite extended BMS$_4$ transformations and local time-dependent
complex Weyl rescalings. Whereas the former corresponds to the
residual symmetry group, the latter represents the natural ambiguity
in the definition of asymptotically flat spacetimes in terms of
conformal compactifications \cite{Penrose1963,Penrose:1965am}.

As a warm-up, we start by re-deriving the known finite transformations
in three dimensions in the asymptotically anti-de Sitter and flat
cases. In the former case, one recovers the Schwarzian derivative as
an application of the AdS$_3$/CFT$_2$ correspondence
\cite{Banados:1999kv,Skenderis:1999nb}. In the latter case, one
obtains the finite transformation laws for the Bondi mass and angular
momentum aspects that have been previously obtained by directly
integrating the infinitesimal transformations
\cite{Barnich:2012rz}. In both these three dimensional cases, these
results are generalized to include local Weyl transformations. In
other words, we are working out the action of finite
Penrose-Brown-Henneaux transformations in the terminology of
\cite{Imbimbo:1999bj,Schwimmer:2000cu}.

Explicit computations are done in the framework of the Newman-Penrose
formalism \cite{Newman1962,Newman1963}, as applied to asymptotically
flat four dimensional spacetimes at null infinity in
\cite{Newman1962a,Exton1969}. Standard reviews are
\cite{newman:1980xx,Chandrasekhar:1985kt,stewart:1991,%
  Stephani:2003tm,Penrose:1986}.

To summarize the results for the simplest case when computations are
done with respect to the Riemann sphere, i.e., when the metric on
$\scrip$ is taken as $d\xbar s^2=0d\widetilde u^2-2d\zeta
d\xbar\zeta$, the extended BMS$_4$ group consists of superrotations
$\zeta=\zeta(\zeta'), \xbar\zeta=\xbar\zeta(\xbar\zeta')$
together with supertranslations $\widetilde
u'=(\ddl{\zeta}{\zeta'}\ddl{\xbar\zeta}{\xbar\zeta'})^{-\half}
[\widetilde u+\beta(\zeta,\xbar\zeta)]$. In particular, the asymptotic
part of the shear, the news, and the Bondi mass aspect 
transform as
\begin{equation}
\sigma'^0_R=(\ddl{\zeta}{\zeta'})^{-\half}(\ddl{\xbar\zeta}{\xbar\zeta'})^{\frac{3}{2}}
  \Big[\sigma^0_R+\xbar\d^2\beta
  +\half\{\xbar\zeta',\xbar\zeta\}(\tilde u+\beta)\Big],
\end{equation}
\begin{equation}
\dot{\sigma}'^0_R=(\ddl{\xbar\zeta}{\xbar\zeta'})^{2}\Big[\dot{\sigma}^0_R
  +\half\{\xbar\zeta',\xbar\zeta\}\Big],
\end{equation}
\begin{multline}
  \label{eq:121}
  (-4\pi G) M'_R=(\ddl{\zeta}{\zeta'}\ddl{\xbar\zeta}{\xbar\zeta'})^{\frac{3}{2}}\Big[
 (-4\pi G) M_R+\xbar\d^2\d^2\beta+\half\{\xbar\zeta',\xbar\zeta\}(\xbar\sigma^0_R+\d^2\beta)
+\\+\half\{\zeta',\zeta\}(\sigma^0_R+\xbar\d^2\beta)
+\frac{1}{4}\{\xbar\zeta',\xbar\zeta\}\{\zeta',\zeta\}(\widetilde u+\beta)
\Big], 
\end{multline}
where $\{\cdot,\cdot\}$ denotes the Schwarzian derivative. 

\section{Adapted Cartan formulation}
\label{sec:adapt-cart-form}

In the Cartan formulation of general relativity, the fundamental
fields are on the one hand, a vielbein, ${e_a}^\mu$, together with its
inverse ${e^a}_\mu$ and associated metric $g_{\mu\nu}={e^a}_\mu
\eta_{ab} {e^b}_\nu$, where $\eta_{ab}$ is constant and, on the other
hand, a Lorentz connection satisfying the metricity condition
$\nabla_a \eta_{bc}=0$,
${\Gamma}_{abc}=\eta_{ad}{\Gamma^d}_{bc}={\Gamma}_{[ab]c}$. Indices
are lowered and raised with $\eta_{ab}$ and $g_{\mu\nu}$ and their
inverses.  The associated connection 1-form is
${\Gamma^a}_b={\Gamma^a}_{bc}e^c$ with $e^c={e^c}_\mu dx^\mu$.  The
torsion and curvature 2-forms are given by $T^a=de^a+{\Gamma^a}_b
\wedge e^b$, ${R^a}_b=d\ {\Gamma^a}_b + {\Gamma^a}_c\wedge
{\Gamma^c}_b$.

Local Lorentz transformations are described by matrices
${\Lambda_a}^b(x)$ with ${\Lambda^a}_c{\Lambda_b}^c=\delta^a_b$. Under
combined frame and coordinate transformations, referred to as gauge
transformations below, the basic variables transform as
\begin{equation}
  \label{eq:1}
\begin{split}
  & {e'_a}^\mu(x')=\Big({\Lambda_a}^b{e_b}^\nu\ddl{{x'}^\mu}{x^\nu}\Big)(x),\\
  & \Gamma'_{abc}(x')=
  \Big( {\Lambda_a}^d\big[{\Lambda_b}^e\Gamma_{def} +
  e_f(\Lambda_{bd})\big]
  {\Lambda_c}^f\Big)(x), 
\end{split}
\end{equation}
where the last expression is equivalent to the transformation law for
the connection 1-form, ${\Gamma^{\prime
    a}}_b={\Lambda^a}_c{\Gamma^{c}}_d{\Lambda_b}^d +{\Lambda^a}_c d
{\Lambda_b}^c $ and $e_a={e_a}^\mu\ddl{}{x^\mu}$.

Equations of motion deriving from the variational principle 
\begin{equation}
  \label{eq:2}
  S[e,\Gamma]=\frac{1}{16\pi G}\int d^d x\, e(
  R_{abcd}\eta^{ac}\eta^{bd}-2\Lambda), 
\end{equation}
are equivalent to $T^a=de^a+{\Gamma^a}_be^b=0$ and Einstein's
equations, $G_{ab}+\Lambda \eta_{ab}=0$. Together with the
metricity condition, the former implies
\begin{equation}
  \label{eq:3}
  \Gamma_{abc}=\half(D_{bac}+D_{cab}-D_{abc}),
\end{equation}
where the structure functions are defined by
${D^c}_{ab}e_c=(e_a({e_b}^\mu)-e_b({e_a}^\mu))\ddl{}{x^\mu}$. Conversely,
$T^a=0$ is equivalent to $D_{cab}=-2\Gamma_{c[ab]}$.

\section{Newman-Penrose formalism in 3d}
\label{sec:newm-penr-form-1}

In three dimensions with $(+,-,-)$ signature, we use
\begin{equation}
  \label{eq:4a}
  \eta_{ab}= \begin{pmatrix} 0 & 1 & 0  \\
1 & 0 & 0  \\0 & 0 & -\frac{1}{2}
\end{pmatrix},
\end{equation}
and the triad $e_a=(l,n,m)$ with associated directional covariant
derivatives denoted by $(D,\Delta,\delta)$. In particular, 
\begin{equation}
g^{\mu\nu}=l^\mu n^\nu+l^\nu n^\mu-2m^\mu m^\nu,\quad \nabla_a= n_a
D+l_a\Delta-2m_a\delta \label{eq:126}. 
\end{equation}
In this case, the spin connection can be dualized,
$\omega_{c\mu}=\frac{1}{4} {\Gamma^{ab}}_\mu\epsilon_{abc}$,
${\Gamma^{ab}}_\mu=\epsilon^{abc}\omega_{c\mu}$ with
$\epsilon_{123}=1$ and $\epsilon^{abc} =
\eta^{ad}\eta^{be}\eta^{cf}\epsilon_{def}$.  The 9 real spin
coefficients are defined by
\begin{equation}
\begin{array}{c|c|c|c}
  \nabla & m^a \nabla l_a &  n^a\nabla l_a & -m^a \nabla n_a \\
  \hline
  D & \kappa=\Gamma_{311}={\omega^2}_1 & 
 \epsilon= \Gamma_{211}=-{\omega^3}_1 & \pi = -\Gamma_{321}={\omega^1}_1
  \\
  \hline
  \Delta & \tau=\Gamma_{312}={\omega^2}_2 & 
\gamma= \Gamma_{212}=-{\omega^3}_2 & \nu = -\Gamma_{322}={\omega^1}_2
  \\
  \hline
  \delta & \sigma=\Gamma_{313} ={\omega^2}_3& 
\beta= \Gamma_{213}=-{\omega^3}_3 & \mu = -\Gamma_{323}={\omega^1}_3
\end{array}
\end{equation}
(see also e.g.~\cite{Milson:2012ry} for slightly different
conventions). It follows that
\begin{equation}
\label{eq:14}
\begin{array}{lll}
 Dl=\epsilon l-2\kappa m, & 
 \Delta l=\gamma l -2\tau m, &
\delta l=\beta l-2\sigma m,\\
 Dn =-\epsilon n+2\pi m, &
\Delta n=-\gamma n+ 2\nu m, &
\delta n=-\beta n+2\mu m,\\
 Dm=\pi l-\kappa n, &
\Delta m=\nu l-\tau n, &
\delta m=\mu l -\sigma n. 
\end{array}
\end{equation}

In order to describe Lorentz transformations, one associates to a real
vector $v=v^a e_a$ a $2\times 2$ symmetric matrix $\hat v=v^{a}\widehat j_{a}$,
where $\widehat j_{a}$ are chosen as
\begin{equation}
\label{generators}
\widehat j_1 = \begin{pmatrix} 1&0\\ 0& 0 \end{pmatrix},\:
\widehat j_2 =\begin{pmatrix} 0 & 0\\ 0&1 \end{pmatrix},\:
\widehat j_3= \frac{1}{2} \begin{pmatrix} 0&1\\ 1&0 \end{pmatrix},
\end{equation}
so that 
\begin{equation}
  \label{invQuad4d1}
  {\rm det}\ \widehat v=\half \eta_{ab}v^av^b, \quad \widehat j_a \epsilon 
  \widehat j_b = \half (\eta_{ab} \epsilon 
  - \epsilon_{abc}\widehat j^c),  \quad \widehat j^a \widehat j_b \widehat j_a =
  \frac{1}{2} \epsilon \widehat j_b \epsilon, 
\end{equation}
where
\begin{equation}
  \epsilon=\begin{pmatrix} 0 & 1\\ -1 & 0
  \end{pmatrix}.
\end{equation}
For $g\in {\rm SL}(2,\mathbb R)$, one considers the
transformation 
\begin{equation}
  g \widehat j_{a} g^{T} v^{a}= \widehat j_a
  {\Lambda^{a}}_{b}v^{b}, \quad  g^T\epsilon= \epsilon g^{-1}.\label{eq:434d}
\end{equation}
More explicitly, if  
\begin{equation}
  \label{eq:74}
  g= \begin{pmatrix}
a & b \\ c & d
\end{pmatrix},
\end{equation}
with $ad-bc=1$ and $a,b,c,d\in \mathbb R$, then
\begin{equation}
\label{Phi}
\Lambda^{a}_{\;\;b}=
\begin{pmatrix}
 a^2 & b^2 & ab \\
 c^2 & d^2 & cd\\
 2 ac & 2bd & ad+bc
\end{pmatrix},\ \Lambda_{a}^{\;\;b}=
\begin{pmatrix}
 d^2 & c^2 & -2 cd\\
b^2 & a^2 & -2 ab \\
- bd & - ac & ad+bc 
\end{pmatrix},
\end{equation}
where the first index is the lign index. $SL(2,\mathbb R)$ group
elements will be parametrized as
\begin{equation}
  \label{eq:19}
  g=\begin{pmatrix} 1 & 0\\ -B & 1 
  \end{pmatrix}
  \begin{pmatrix}
    1 & -A \\ 0 & 1
  \end{pmatrix}
\begin{pmatrix}
    e^{-E/2} & 0 \\ 0 & e^{E/2}
  \end{pmatrix}
=\begin{pmatrix}
    e^{-E/2}  &  -Ae^{E/2} \\  -Be^{-E/2} & (1+AB)e^{E/2}
  \end{pmatrix}.
\end{equation}
Using $\omega^a=\half\epsilon^{abc}\Gamma_{bc}$ and the transformation
law of the Lorentz connection given below \eqref{eq:1}, we have
\begin{equation}
  \label{eq:15}
  \omega^{\prime a}={\Lambda^a}_b\omega^b+\half \epsilon^{abc}{\Lambda_b}^d
  d\Lambda_{cd}. 
\end{equation}
In terms of $\hat\omega=\hat j_a\omega^a$, this is equivalent to
\begin{equation}
  \label{eq:18}
  \hat\omega'= g\hat\omega g^T- g \epsilon dg^T
\end{equation}
Explicitly, for the spin coefficients encoded in 
\begin{equation}
  \label{eq:43}
  \hat\omega_1=\begin{pmatrix} \pi & -\half\epsilon\\ -\half\epsilon & 
    \kappa\end{pmatrix},\quad \hat\omega_2=\begin{pmatrix} \nu &
    -\half\gamma\\ -\half\gamma &  
    \tau \end{pmatrix}, \quad \hat\omega_3=\begin{pmatrix} \mu &
    -\half\beta\\ -\half\beta & 
    \sigma\end{pmatrix},
\end{equation}
one finds
\begin{equation}
  \label{eq:13}
  \hat\omega'_a={\Lambda_a}^c g\hat\omega_c g^T- g\epsilon
  e'_a(g^T). 
\end{equation}

In this case, Einstein's equations are equivalent to 
\begin{equation}
  \label{eq:24}
  d\hat e-2\hat\omega\epsilon\hat e=0, \quad
  d\hat \omega-\hat\omega\epsilon\hat\omega-\frac{\Lambda}{2}\hat
  e\epsilon\hat e=0. 
\end{equation}

Alternatively, one can use $\widecheck v = \hat v \epsilon$ in order
to describe real vectors by traceless $2 \times 2$ matrices. The
associated basis is
\begin{equation}
\label{generatorsa}
\widecheck j_1 = \begin{pmatrix} 0&1\\ 0& 0 \end{pmatrix},\:
\widecheck j_2 =\begin{pmatrix} 0 & 0\\ -1& 0 \end{pmatrix},\:
\widecheck j_3= \frac{1}{2} \begin{pmatrix} -1&0\\ 0&1 \end{pmatrix},
\end{equation}
so that 
\begin{equation}
\label{invQuad4d1a}
{\rm tr}\ \widecheck v^2= \eta_{ab}v^av^b,\quad \widecheck j_a
\widecheck j_b = -\half (\eta_{ab}  
+ \epsilon_{abc}\widecheck j^c),  \quad \widecheck j^a \widecheck
j_b \widecheck j_a = \frac{1}{2} \widecheck j_b. 
\end{equation}
In this case, we have
\begin{gather}
  g \widecheck j_{a} g^{-1} v^{a}= \widecheck j_a
  {\Lambda^{a}}_{b}v^{b},\label{eq:434da}\quad
\widecheck \omega' = g \widecheck \omega g^{-1} -d g g^{-1},
\quad \widecheck\omega'_a={\Lambda_a}^c g\widecheck\omega_c g^{-1}-
e'_a(g) g^{-1},
\end{gather}
where
\begin{equation}
  \label{eq:43a}
  \widecheck\omega_1=\begin{pmatrix}\half\epsilon & \pi \\ -\kappa &
    -\half\epsilon  
    \end{pmatrix},\quad \widecheck\omega_2=\begin{pmatrix} \half \gamma &
      \nu \\ -\tau & -\half\gamma \end{pmatrix}, 
\quad \widecheck\omega_3=\begin{pmatrix} \half \beta & \mu\\ -\sigma &
  -\half\beta\end{pmatrix}, 
\end{equation}
and
\begin{equation}
  \label{eq:24a}
  d\widecheck e-2\widecheck\omega\widecheck e=0, \quad 
d\widecheck
\omega-\widecheck\omega\widecheck\omega-\frac{\Lambda}{2}\widecheck
e\widecheck e=0. 
\end{equation}

\section{3d asymptotically AdS spacetimes at spatial infinity}
\label{sec:3d-asympt-spac}

\subsection{Fefferman-Graham solution space}
\label{sec:feff-grah-solut}

In the AdS$_3$ case, $\Lambda=-L^{-2}\neq 0$, we start by rederiving
the general solution to the equations of motion in the context of the
Newman-Penrose formalism. We will recover the on-shell bulk metric of
\cite{Banados:1998sm}, but with an arbitrary conformal factor for the
boundary metric \cite{Skenderis:1999nb} (see also Section 2 of
\cite{Barnich:2010eb} in the current context).

The analog of the Fefferman-Graham gauge
fixing is to assume that 
\begin{equation}
	\mu = \beta = \sigma = 0.
\end{equation}
which is equivalent to $\Gamma_{ab3}=0$ and can be achieved by a local
Lorentz transformation. This means that the triad is parallely
transported along $m$ and that $m$ is the generator of an affinely
parametrized spatial geodesic. In this case, $\nabla_{[a}
m_{b]}=n_{[a}l_{b]}(\pi+\tau)$ so that $m$ is hypersurface
orthonormal if and only it is a gradient, which in turn is equivalent
to 
\begin{equation}
\pi=-\tau\label{eq:11}. 
\end{equation}
This condition will also be imposed in the following.

Introducing coordinates $x^\mu = (x^+,x^-, \rho)$, $\mu = 1,2,3$ such
that $m$ is normal to the surfaces $\rho =cte$ and the coordinate
$\rho$ is the suitably normalized affine parameter on the geodesic
generated by $m$, the triad takes the form
\begin{equation}
m=\frac{\d}{\d \rho},\quad l = l^a\frac{\d}{\d x^a}, \quad n =
	n^a\frac{\d}{\d x^a}.
\end{equation}
where $a=(+,-)$. The associated cotriad is
\begin{equation}
	e^1 = \frac{\epsilon_{ab}n^b dx^a}{e}, \quad e^2 =
        \frac{\epsilon_{ab}l^a dx^b}{e}, \quad e^3 = d\rho, \quad
e=\epsilon_{ab}l^a n^b, 
\end{equation}
where $\epsilon_{+-}=1=-\epsilon_{-+}$ and $\epsilon_{\pm\pm}=0$. 
In order to compare with the general solution given in
\cite{Barnich:2010eb}, one introduces an alternative radial
coordinate $r = e^{\frac{\rho}{\sqrt 2 L}}$, in terms of which 
\begin{equation}
	m = \frac{r}{\sqrt 2 L}\frac{\d}{\d r}, \quad e^3 = \sqrt 2 \frac{L}{r} dr.
\end{equation}

Under these assumptions, the Newman-Penrose field equations
\eqref{eq:NPeq3DII}-\eqref{eq:NPeq3DXII} can be solved exactly.
Indeed, the three equations \eqref{eq:NPeq3DII}, \eqref{eq:NPeq3DIX}
and \eqref{eq:NPeq3DVII} reduce to the system
\begin{equation}
	\delta \kappa  =  2 \tau \kappa, \quad
	\delta \nu  =  2 \tau \nu,\quad
	\delta \tau  =  \tau^2 - \kappa\nu - \frac{1}{2L^2},
\end{equation}
which is solved by introducing the complex combinations $\cL_{\pm} =
\tau \pm i \sqrt {\nu \kappa}$. The general solution is 
given by
\begin{equation}
\begin{split}
 & \tau = - \pi = \frac{-1}{\sqrt 2 L k} (r^4 - C_1^2 + C_2C_3), \quad
  \kappa = \frac{-\sqrt{2}C_2 r^2}{ Lk} , \quad \nu =
  \frac{\sqrt{2}C_3 r^2}{ L k},\\
& k = r^4 - 2 C_1 r^2+ C_1^2 - C_2C_3. \label{eq:adstetradsol1}
\end{split}
\end{equation}
The last two radial equations involving the
spin coefficients, equations \eqref{eq:NPeq3DIV} and
\eqref{eq:NPeq3DVIII}, simplify to
\begin{equation}
	\label{eq:ADSsysI}
	\delta \epsilon  = \tau \epsilon + \kappa \gamma,\quad
	\delta \gamma = \tau\gamma - \nu \epsilon,
\end{equation}
and are solved through 
\begin{equation}
	\epsilon = C_4 \frac{r^3 - C_1 r}{k} +C_5\frac{C_2 r}{k},\quad
	\gamma  =  C_5 \frac{r^3 - C_1 r}{k} + C_4 \frac{C_3 r}{k}.
	\label{eq:adstetradsol2}
\end{equation}
The last radial
equations are \eqref{eq:NPeq3DXI} and \eqref{eq:NPeq3DXII}. Their
$r$-component are trivially satisfied while their components along
$x^\pm$ are of the same form than \eqref{eq:ADSsysI},
\begin{equation}
	\delta l^\pm  = \tau l^\pm + \kappa n^\pm,\quad
	\delta n^\pm  = \tau n^\pm - \nu l^\pm,
\end{equation}
which leads to 
\begin{equation}l^\pm = K^\pm_1 \frac{r^3 - C_1 r}{k} + K^\pm_2
  \frac{C_2 r}{k},\quad 
	n^\pm =  K^\pm_2 \frac{r^3 - C_1 r}{k} + K^\pm_1 \frac{C_3
          r}{k}.\label{eq:adstetradsol3} 
\end{equation}
In these equations, $C_i$, $K_1^\pm$, $K_2^\pm$ are functions of $x^a
= x^\pm$ alone.

Note that asymptotic invertibility of
the triad is controlled by the invertibility of the
matrix formed by these functions,
\begin{equation}
	\label{eq:matrixK}
	\left(\begin{array}{cc} K_1^+ & K_2^+ \\ K_1^- & K_2^-
\end{array} \right), \quad \epsilon_{ab}K_1^aK_2^b\neq 0. 
\end{equation}
Using the radial form of the various quantities, equations
\eqref{eq:NPeq3DI} and \eqref{eq:NPeq3DV} are equivalent to
\begin{equation}
\begin{split}
	\label{eq:adsc2c3I}
	K^a_1 \d_a C_1 - K_2^a \d_a C_2 + 2 C_2 C_5  =  0,\\
	K^a_1 \d_a C_3 - K_2^a \d_a C_1 + 2 C_3 C_4  =  0,
\end{split}
\end{equation}
which then implies that equation \eqref{eq:NPeq3DIII} reduces to
\begin{equation}
	\label{eq:adsc1} K_1^a\d_a C_5 - K_2^a\d_a C_4 + 2 C_4C_5 +
\frac{4}{L^2}C_1=0,
\end{equation}
while the components along $x^\pm$ of equation \eqref{eq:NPeq3DX}
become 
\begin{equation}
	\label{eq:adsc5c4}
	K_1^a\d_a K_2^\pm - K_2^a\d_a K_1^\pm +
        C_5K_1^\pm + C_4K_2^\pm=0. 
\end{equation}
Because of invertibility of the matrix \eqref{eq:matrixK}, equations
\eqref{eq:adsc5c4} and \eqref{eq:adsc1} can be used to express
$C_4,C_5$ and $C_1$ in terms of $K_1^a$ and $K_2^a$. The two equations
in \eqref{eq:adsc2c3I} then become dynamical equations for $C_2$ and
$C_3$. Since we now have treated all Newman-Penrose equations, the
solution space is parametrized by $K^a_1,K^a_2$ and by initial
conditions for $C_2$ and $C_3$.

In the limit $r$ going to infinity, the triad elements $l$ and $n$
given in \eqref{eq:adstetradsol3} take the form $l^\pm = r^{-1}
K_1^\pm + O(r^{-3})$, $n^\pm = r^{-1} K_2^\pm + O(r^{-3})$.  With a
change of coordinates on the cylinder, we can make the associated
asymptotic metric explicitly conformally flat. This amounts to the
choice
\begin{equation}
  \label{eq:adsassnl} K_1^+ = 0, \quad
  K_1^- = \sqrt 2 e^{-\varphi}, \quad K_2^+ = \sqrt 2 e^{-\varphi},
  \quad K_2^- = 0.
\end{equation} 
Introducing this into equations \eqref{eq:adsc1}
and \eqref{eq:adsc5c4}, we get
\begin{equation}
	C_4 = \sqrt 2 e^{-\varphi} \d_- \varphi, \quad C_5 = - \sqrt 2
	e^{-\varphi} \d_+ \varphi,\quad  C_1 = L^2 e^{-2\varphi}
        \d_-\d_+ \varphi, 
\end{equation}
while the dynamical equations \eqref{eq:adsc2c3I} reduce to
\begin{equation}
\begin{split}
	\label{eq:adsc2phi}
	& \d_+C_2 + 2 \d_+ \varphi C_2 = L^2 \d_- \left( e^{-2\varphi}
          \d_-\d_+
          \varphi\right),\\
	& \d_-C_3 + 2 \d_- \varphi C_3 = L^2 \d_+ \left( e^{-2\varphi}
          \d_-\d_+ \varphi\right).
\end{split}
\end{equation}

With the extra conditions \eqref{eq:adsassnl}, the space of solutions
is parametrised by three functions $\varphi, C_2$ and $C_3$ defined on
the cylinder with coordinates $x^\pm$ such that equations
\eqref{eq:adsc2phi} are valid.  These two equations can be integrated
directly but we will derive the explicit form of $C_2$ and $C_3$ in a
different way using the action of the asymptotic symmetry group below.

\subsection{Residual gauge symmetries}
\label{sec:resid-gauge-symm-1}

The residual gauge transformations are the finite gauge
transformations that preserve the set of asymptotic solutions. Since
these transformations map solutions to solutions, once the conditions
that determine the asymptotic solution space are preserved, no further
restrictions can arise. A gauge transformation is a combination of a
local Lorentz transformation and a change of coordinates of the form
\begin{equation}
	r = r(r', x'^\pm), \quad x^\pm = x^\pm(r', x'^\pm).
\end{equation}
The unknowns are $A, B, E, r$ and $x^\pm$ as functions of $r', x'^\pm$.

Using the $a=3$ component of the transformation law of the triad,
\begin{equation}
	e'^\mu_a \frac{\d x^\nu}{\d x'^\mu} = \Lambda_a^{\phantom a b}
	e_b^\nu,
\end{equation}
the requirement $m'^\mu = \frac{r'}{\sqrt 2 L}\delta_{r'}^{\mu}$ is
equivalent to 
\begin{equation}
  \frac{r'}{\sqrt 2 L}\d_{r'} x^\mu =
  -bd l^\mu - ac n^\mu + (ad+bc) m^\mu.
\end{equation}
Expanding for each coordinate, we get
\begin{equation}
\begin{split}
	\label{eq:adsnormalchangeI}
	\frac{r'}{\sqrt 2 L}\frac{\d x^+}{\d r'} & = \frac{\sqrt 2 r
        }{k} e^{-\varphi}\left( A(1+AB)e^EC_2 + B e^{-E}
          (r^2 - C_1)\right),\\
	\frac{r'}{\sqrt 2 L}\frac{\d x^-}{\d r'} & = \frac{\sqrt 2
          r}{k} e^{-\varphi}\left( A(1+AB)e^E (r^2 - C_1) + B e^{-E}
          C_3\right),\\
	\frac{r'}{\sqrt 2 L}\frac{\d r}{\d r'} & = (1+2AB)
        \frac{r}{\sqrt 2 L}.
\end{split}
\end{equation}
In order to implement the gauge fixing condition on the new spin
coefficients $\widecheck \omega'_3 = 0$, we first rewrite the last
equation of \eqref{eq:434da} as
\begin{equation}
	g^{-1}e'_a(g)={\Lambda_a}^b \widecheck\omega_b-g^{-1}\widecheck\omega'_ag.
\end{equation}
When $a=3$ this becomes
\begin{equation}
	g^{-1} \delta' g = \Lambda_3^{\phantom 1 b} \widecheck \omega_b,
\end{equation}
and is equivalent to three conditions on the Lorentz parameters, 
\begin{equation}
	d\delta'a - b\delta'c = \Lambda_3^{\phantom 1 b}(\widecheck \omega_b)_{11}, \
	d\delta'b - b\delta'd = \Lambda_3^{\phantom 1 b}(\widecheck \omega_b)_{12}, \
	a\delta'c - c\delta'a = \Lambda_3^{\phantom 1 b}(\widecheck \omega_b)_{21}.
\end{equation}
When suitably combining these equations, one finds
\begin{equation}
\begin{split}
	\label{eq:adsnormalchangeIV} \frac{r'}{\sqrt 2 L}\frac{\d
B}{\d r'} & = A(1+AB) e^{2E} \kappa + B \tau,\\ \frac{r'}{\sqrt 2
L}\frac{\d A}{\d r'} & = -A(1+AB) \pi - B e^{-2E} \nu + A^3 (1+AB)
e^{2E} \kappa + A^2 B \tau,\\ \frac{r'}{\sqrt 2 L}\frac{\d E}{\d r'} &
= A(1+AB) e^E \epsilon - B e^{-E} \gamma - 2 A^2 (1+AB)e^{2E}\kappa -
2AB \tau.
\end{split}
\end{equation}
  
The set of equations \eqref{eq:adsnormalchangeI} and
\eqref{eq:adsnormalchangeIV} forms a system of differential equations
for the radial dependence of the unknown functions. In order to solve
it asymptotically, we will assume that the functions have the
following asymptotic behavior, 
\begin{equation}
	r = O(r'), \quad x^\pm, E = O(1), \quad A, B = O(r'^{-1}).
\end{equation}
Inserting this into the equations, we easily get
\begin{equation}
\begin{split} 
&B = B_0(x'^\pm) r'^{-1} + O(r'^{-3}), \quad A =
A_0(x'^\pm) r'^{-1} + O(r'^{-3}), \\ & E = E_0(x'^\pm) - (Le^{-\varphi}
\d_- \varphi A_0 + Le^{-\varphi} \d_+ \varphi B_0 + A_0 B_0)r'^{-2} +
O(r'^{-4}),\\ 
& r = e^{r_0(x'^\pm)} r' - A_0 B_0 e^{r_0} r'^{-1} +
O(r'^{-3}),\\ 
& x^+ = x_0^+(x'^\pm) - L B_0 e^{-\varphi - E_0 - r_0}
r'^{-2} + O(r'^{-4}), \\ 
& x^- = x_0^-(x'^\pm) - L A_0 e^{-\varphi + E_0
- r_0} r'^{-2} + O(r'^{-4}),
\end{split}
\end{equation} 
where we have assumed $\frac{r}{r'} >0$ asymptotically. At this stage,
we have fixed the radial dependence of all the unknown functions and
we are left with six functions $A_0, B_0, E_0, r_0,x_0^\pm$ of $x'^\pm$. 

We now have to require $m'_\mu = \frac{\sqrt 2 L}{r'}\delta^{r'}_\mu$.
However, since we have inmposed $m' = \frac{r'}{\sqrt 2 L}
\ddl{}{r'}$, we already have $m'_{r'} = \frac{\sqrt 2 L}{r'}$. This
follows from $m'_{r'}={\Lambda^3}_b{e^b}_\nu\ddl{x^\nu}{r'}$ on the
one hand and from $\frac{r'}{\sqrt 2
  L}\ddl{x^\nu}{r'}={\Lambda_3}^c{e_c}^\nu$ on the other. For the
remaining components of $m'_\mu$ it is enough to verify that
$m'_{\pm'}=o({r'}^0)$ since solutions are transformed into solutions
under local Lorentz and coordinate transformations. Indeed,
$de'^{a}+{\Gamma'^{a}}_be'^{b}=0$, and for $a=3$, $de'_3
+\Gamma'_{3b}e^{'b}=0$. Contracting with ${e'_3}^\mu$ then implies
that $\frac{r'}{\sqrt 2 L}\d_{r'}m_{\nu'}-\frac{r'}{\sqrt 2
  L}\d_{\nu'}\frac{\sqrt 2 L}{r'}
+\Gamma'_{3b3}{e^{'b}}_{\nu'}-\Gamma'_{33\nu'}=0$, which reduces to
$\d_{r'} m'_\nu = -\frac{\sqrt 2 L}{r'^{2}}
\delta^{r'}_\nu$. Extracting the leading order from
\begin{equation}
	\label{eq:adseprim3}
	e'^3_\mu = 2ac \,e^1_\nu \frac{\d x^\nu}{\d x'^\pm} + 2bd\,
        e^2_\nu \frac{\d x^\nu}{\d x'^\pm} 
	 + (ad + bc) e^3_\nu \frac{\d x^\nu}{\d x'^\pm},
\end{equation}
we then get
\begin{equation}
	\label{eq:adsAB}
	B_0 e^{-E_0 + r_0 + \varphi} \frac{\d x^-_0}{\d x'^\pm} + 
	A_0 e^{E_0 + r_0 + \varphi} \frac{\d x^+_0}{\d x'^\pm} = L \frac{\d
	r_0}{\d x'^\pm}.
\end{equation}

The last condition we have to require is the asymptotically
conformally flat form of the new triad. This can be done by imposing
\begin{equation}
	e'^1 = \frac{r'e^{\varphi'}}{\sqrt 2} d x'^- + O (r'^{-1}),
		\quad 
		e'^2 = \frac{r' e^{\varphi'}}{\sqrt 2} d x'^++ O (r'^{-1}).
\end{equation}
The leading terms of $e'^1_\nu = \Lambda^1_{\phantom 1 a} e^a_\mu \frac{\d
x^\mu}{\d x'^\nu}$ and $e'^2_\nu = \Lambda^2_{\phantom 1 a} e^a_\mu \frac{\d
x^\mu}{\d x'^\nu}$ yield
\begin{equation}
	\label{eq:adsgaugeconst}
	\frac{\d x_0^+}{\d x'^-} = 0= 
	\frac{\d x_0^-}{\d x'^+} = 0, \quad
	e^{-E_0 + r_0 + \varphi} \frac{\d x_0^-}{\d x'^-}=e^{\varphi'}
        = e^{E_0 + r_0 + \varphi} \frac{\d x_0^+}{\d x'^+}. 
\end{equation}
Combining with equation \eqref{eq:adsAB}, allows one to extract $A_0,
B_0,E_0$ in terms of the other functions, 
\begin{equation}
	E_0 = \frac{1}{2} \ln \frac{\d'_- x_0^-}{\d'_+
	x_0^+}, \quad A_0 = L
	\frac{e^{-\varphi-r_0}}{\sqrt{ \d'_- x_0^- \d'_+
	x_0^+}}\frac{\d
	r_0}{\d x'^+},\quad B_0 = L \frac{e^{-\varphi-r_0}}{\sqrt{ \d'_- x_0^- \d'_+
	x_0^+}} \frac{\d
	r_0}{\d x'^-}.
\end{equation}
It thus follows that the residual gauge symmetries are determined (i)
by the change of variables $x^\pm=x^\pm_0(x^{\prime\pm})$ at infinity,
each depending on a single variable, which we assume to be orientation
preserving $\d'_+x^+_0>0<\d'_-x^-_0$, and (ii) by $r_0(x'^+,x'^-)$.

For notational simplicity, we drop the subscript $0$ on the change of
variables at infinity and on the Weyl parameter in the next section. 

\subsection{Action of conformal and Weyl group}
\label{sec:penr-brown-henn}

The group obtained in the previous section is
the combined conformal and Weyl group and is parametrized by 
\begin{equation}
\Big(x^{\prime +}(x^+),x^{\prime
  -}(x^-),r(x^{\prime+},x^{\prime-})\Big).\label{eq:31}
\end{equation}
The last equation of \eqref{eq:adsgaugeconst} encodes the transformation
law of $\varphi$, 
\begin{equation}
	\label{eq:adstransphi}
	\varphi'(x'^+,x'^-) = \varphi(x^+,x^-) +
        r(x'^+,x'^-) + \frac{1}{2} \ln \left(\d'_+ x^+ \d'_- 
          x^-\right). 
\end{equation}
Note that, as a consequence, if $r(x'^+,x'^-), r^s(x''^+,x''^-)$ and
$r^c(x''^+,x''^-)$ are associated to a first, a second successive and the
combined transformation respectively, the composition law is
\begin{equation}
r^c(x''^+,x''^-)=r^s(x''^+,x''^-) + r(x'^+,x'^-).
\end{equation}

This group reduces to the conformal group
for fixed conformal factor of the boundary metric: when
$\varphi=\varphi'$ it follows from (\ref{eq:adstransphi}) that $r$ is
determined by the change of variables at infinity,
$r=-\half\ln{(\d'_+x^+\d'_-x^-)}$. When freezing the coordinate
transformations, one remains with the additive group of Weyl
rescalings that amount here to arbitrary shifts of $\varphi$.

As discussed in Section \ref{sec:feff-grah-solut}, the on-shell
metric, triads and spin connections are entirely determined by the
arbitrary conformal factor $\varphi(x^+,x^-)$ and the integration
functions $C_2(x^+,x^-),C_3(x^+,x^-)$ satisfying \eqref{eq:adsc2phi}.
To obtain the action of the group on the latter, we can extract the
subleading terms of $l'( x^+) = \Lambda_1^{\phantom 1b}e_b^+$ and $n'(
x^-) = \Lambda_2^{\phantom 1b}e_b^-$. This gives 
\begin{equation}
\begin{split}
	\label{eq:ads3transC2}
	C'_2 & =  e^{-2r} \frac{\d'_-x^-}{\d'_+x^+} C_2 + L^2
	e^{-2\varphi'} \left(\d'^2_- r + (\d'_-r)^2 - 2 \d'_-r
	\d'_-\varphi'\right),\\
	C'_3 & =  e^{-2r} \frac{\d'_+x^+}{\d'_-x^-} C_3 + L^2
	e^{-2\varphi'} \left(\d'^2_+ r + (\d'_+r)^2 - 2 \d'_+r
	\d'_+\varphi'\right),
\end{split}
\end{equation}
which can also be written in terms of $\varphi$ using equation
\eqref{eq:adstransphi}. Note that, by construction, the transformed
$C'_2(x'^+,x'^-),C_3(x'^+,x'^-)$ have to satisfy the transformed
equations, i.e., equations \eqref{eq:adsc2phi} where all quantities,
$C_2,C_3,\varphi$, $x^\pm,\d_\pm$ 
are primed. 

In the particular case where
$\varphi=0$, equations \eqref{eq:adsc2phi} reduce to $\d_+ C_{2R} =
0$, $\d_- C_{3R}=0$ so that $C_{2R}=
(8\pi G L) T_{--}(x^-)$ and $C_{3R} = (8\pi G L)
T_{++}(x^+)$. Applying the particular Weyl transformation $x'^\pm =
x^\pm$, $r = \varphi'$, and removing all primes, we obtain from
\eqref{eq:ads3transC2} that the general solution to the dynamical
equations \eqref{eq:adsc2phi} for arbitrary $\varphi$ is given by 
\begin{equation}
\begin{split}
	C_2  &=  e^{-2\varphi}L^2\left[\frac{8 \pi G}{L} T_{--}(x^-) + 
	\d^2_- \varphi - (\d_-\varphi)^2\right], \\ 
	C_3 & =  e^{-2\varphi}L^2\left[ \frac{8 \pi G}{L}T_{++}(x^+) + 
	\d^2_+ \varphi - (\d_+\varphi)^2\right].
\end{split}
\end{equation}

Solution space can thus also be parametrized by the conformal factor
$\varphi$ and the two integration functions $T_{\pm\pm}(x^\pm)$
depending on a single variable each. The action of the asymptotic
symmetry group on the latter can be extracted from equations
\eqref{eq:ads3transC2},
\begin{equation}
  \label{eq:55}
\begin{split}
&  T'_{\pm\pm}(x^{\prime\pm})=(\d'_\pm
  x^\pm)^2T_{\pm\pm}(x^\pm_0)-\frac{c_\pm}{24\pi}\{x^\pm,x^{\prime\pm}\},\\
& \iff T'_{\pm\pm}(x^{\prime\pm})=(\d_\pm
  x'^\pm)^{-2}\Big[T_{\pm\pm}(x^\pm)+\frac{c_\pm}{24\pi}\{x^{\prime\pm},x^\pm\}\Big], 
\quad
  c_\pm=\frac{3L}{2G}, 
\end{split}
\end{equation}
in terms of the Schwarzian derivative for a function $F$ of $x$, 
\begin{equation}
\{F,x\}=\d^2_x\ln\d_xF-\half (\d_x\ln
\d_xF)^2,
\end{equation}
and with the characteristic values of the central charges for
asymptotically AdS$_3$ gravity~\cite{Brown:1986nw}.  In other words,
the integration functions $T_{\pm\pm}$ are Weyl invariant, while under
the centrally extended conformal group, one recovers the well-known
coadjoint action, i.e., the standard transformation law of an
energy-momentum tensor.

\section{3d asymptotically flat spacetimes at null infinity}
\label{sec:3d-asympt-flat}

\subsection{Solution space}
\label{sec:newman-unti-solution}

The first gauge fixing conditions that we will assume are 
\begin{equation}
  \label{eq:18a}
  \kappa=\epsilon=\pi=0.
\end{equation}
This is equivalent to $\Gamma_{ab1} = 0$ which can be achieved by a
suitable Lorentz rotation. It implies that the tetrad is parallely
transported along $l$ and that $l$ is the generator of an affinely
parametrized null geodesic. In this case, $\nabla_{[a}
l_{b]}=-2l_{[a}m_{b]}(\tau-\beta)$, so that $l$ is always hypersurface
orthornormal. It is a gradient if and only if 
\begin{equation}
\tau=\beta,\label{eq:9}
\end{equation}
a condition which will also be imposed in the following.

Introducing Bondi coordinates $x^\mu=(u,r,\phi), \mu = 0,1,2$ such
that the surfaces $u=cte$ are null with normal vector $l$, $l_\mu =
\delta^0_\mu$ and such that $r$ is the suitably normalized affine
parameter on the null geodesics generated by $l$, the triad takes the
form
\begin{equation}
  \label{eq:23}
l= \frac{\d}{\d r},\quad  n=\ddl{}{u}+W\ddl{}{r}+V\ddl{}{\phi}, \quad
   m=U\ddl{}{r}+T\ddl{}{\phi}.
 \end{equation}
The associated cotriad is 
\begin{equation}
	e^1=(-W+T^{-1}UV)du+dr- T^{-1}Ud\phi, \quad e^2=du \quad     
    	e^3=-T^{-1}Vdu+T^{-1}d\phi.\label{eq:29}
\end{equation}

Under these assumptions, the Newman-Penrose equations
\eqref{eq:NPeq3DII}-\eqref{eq:NPeq3DV} fix the $r$
dependence of all spin coefficients according to
\begin{equation}
  \label{eq:27}
\begin{split}
 & \sigma=-\frac{1}{2}\frac{1}{r+C_1},\quad 
  \tau=\frac{C_2}{r+C_1}=\beta,\\ & \gamma= C_3
  -\frac{2C_2^2}{r+C_1}, \quad \mu= \frac{C_4}{r+C_1} , \quad \nu= C_5 -
  \frac{2C_2C_4}{r+C_1}.
\end{split}
\end{equation}
for $C_i=C_i(u,\phi)$. When used in equations \eqref{eq:NPeq3DX} and
\eqref{eq:NPeq3DXI}, the $r$ dependence of the triad is
\begin{equation}
    \label{eq:28}
\begin{split}
& T=\frac{K_1}{r+C_1},\quad V=-\frac{2K_1C_2}{r+C_1}+K_2,\quad
  U=-C_2+\frac{K_3}{r+C_1},\\
& W = -C_3r + K_4 -
\frac{2C_2K_3}{r+C_1}, 
\end{split}
\end{equation}
with $K_a=K_a(u,\phi)$.

In order to solve the remaining equations, we will assume in addition
that 
\begin{equation}
	\sigma = -\frac{1}{2r} + O(r^{-3}), \quad \tau = O(r^{-2}), \quad V =
	O(r^{-1}).
\end{equation}
The first condition can be satsified by changing the affine parameter
$r \rightarrow r+C_1$. We can then do a Lorentz transformation with
$a=d=1$, $c=0$ and $b=C_2$ in order to impose $C_2=0$, and finally a
change of coordinates $\d_u\phi'=-K_2\d_\phi \phi'$ to obtain $K_2=0$.
Note however that both of these last two transformations are only
valid asymptotically. Requiring them to preserve the gauge fixing
conditions will require subleading terms in a similar way as in the
computation of section \ref{sec:resid-gauge-symm}. On the level of
solutions, the additional conditions simply amount to setting
\begin{equation}
  \label{eq:16}
  C_1=C_2=K_2=0. 
\end{equation}

Redefining $K_1=e^{-\varphi}$, the remaining equations, i.e.,
\eqref{eq:NPeq3DIX}-\eqref{eq:NPeq3DVII} and \eqref{eq:NPeq3DXII},
are equivalent to 
\begin{equation}
  C_4=\half K_4,\quad C_3=\d_u\varphi,\quad 
  C_5= e^{-\varphi} \d_u\d_\phi\varphi,
\end{equation}
where
\begin{equation}
\begin{split}
	\d_u K_4 + 2 \d_u \varphi K_4 &=
	2 e^{-2\varphi} (\d_u\d_\phi^2\varphi - \d_\phi \varphi
	\d_u\d_\phi \varphi),\label{eq:bms3evolI}\\ 
  \d_u K_3 + 2 \d_u \varphi K_3 &= e^{-\varphi} \d_\phi K_4.
\end{split}
\end{equation}
These equations can be integrated directly, but we will again generate
the solution by using the asymptotic symmetry group below. 

In this case, \eqref{eq:27} and \eqref{eq:28} simplify to 
\begin{equation}
\begin{split}
  & \sigma = -\frac{1}{2r}, \quad \tau = \beta = 0, \quad \gamma =
  \d_u \varphi,\quad \mu = \frac{K_4}{2r}, \quad \nu =
  e^{-\varphi} \d_u\d_\phi \varphi,\\
  & T=\frac{e^{-\varphi}}{r}, \quad W = - \d_u\varphi r + K_4, \quad U =
  \frac{K_3}{r}, \quad V = 0.
\end{split}
\end{equation}

\subsection{Residual gauge symmetries}
\label{sec:resid-gauge-symm}

The residual gauge symmetries again consist of the subset of gauge
transformations that preserve the set of conditions determining
the asymptotic solution space.  We will consider a general change of
coordinates of the form
\begin{equation}
u  =  u(u', r', \phi'),\quad r  =  r(u', r', \phi'),\quad \phi  =
\phi(u', r', \phi'),\label{coc}  
\end{equation}
combined with an arbitrary local Lorentz transformation. The unknowns
are $A, B, E, u, r, \phi$  
as functions of $u', r', \phi'$. 

Using the $a=1$ component of the transformation law for the triad
\begin{equation}
	e^{\prime\mu}_a \frac{\d x^\nu}{\d x^{\prime \mu}} =
	\Lambda_a^{\phantom a b} {e_b}^{\nu},
\end{equation}
it follows that imposing $l^{\prime\mu} = \delta_{r'}^\mu$ is equivalent to
the radial equations, 
\begin{equation}
\begin{split}
	\label{eq:3dradialchI}
	\frac{\d u}{\d r'} & =  B^2 e^{-E},\\
	\frac{\d \phi}{\d r'} & =  2B(1+AB) T ,\\
	\frac{\d r}{\d r'} & =  (1+AB)^2 e^{E} + B^2 e^{-E} W + 2 B
        (1+AB) U.
\end{split}
\end{equation}
The gauge fixing on the new spin coefficients takes the form $\widecheck\omega'_1
= 0$. The component $a=1$ of the last equation of \eqref{eq:434da}  can be rewritten as
\begin{equation}
	g^{-1} \d_{r'} g = \Lambda_1^{\phantom 1 b} \widecheck \omega_b.
\end{equation}
This is equivalent to three conditions on the rotation parameters, which can
be suitably combined to yield
\begin{equation}
\begin{split}
	\label{eq:3dradialchIV}
	\frac{\d B}{\d r'} & =  2B(1+AB) e^{E}\sigma,\\
	\frac{\d A}{\d r'} & =  - B^2 e^{-2E} \nu - 2 B(1+AB)e^{-E} \mu + 2A^2 B(1+AB) e^{E}\sigma,\\
	\frac{\d E}{\d r'} & = - B^2 e^{-E} \gamma-  4AB(1+AB)
        e^{E}\sigma.
\end{split}
\end{equation}

The set of equations \eqref{eq:3dradialchI} and
\eqref{eq:3dradialchIV} forms a system of differential equations for
the radial dependence of the unknown functions. In order to solve it
asymptotically, we assume that the functions have the following
asymptotic behavior,
\begin{equation}
	r= O(r'), \quad   A, E, u, \phi = O(1), \quad B = O(r'^{-1}).
\end{equation}
The unknown $r$ can be traded for $\chi =
re^{-E} = O(r')$ satisfying
\begin{equation}
	\frac{\d \chi}{\d r'} = 1 - A^2B^2 + B^2 e^{-2E} K_4 +
	\frac{2B}{\chi}(1+AB) e^{-2E} K_3 = 1 + O(r'^{-2}).
\end{equation}
The solution is given by
\begin{equation}
	\chi = r' + \chi_0(u, \phi) + O(r'^{-1}),
\end{equation}
which, when introduced into the other radial equations, gives
\begin{eqnarray*}
	B & = & B_0(u, \phi)r'^{-1} + (A_0 B_0^2 - B_0 \chi_0) r'^{-2} +
	O(r'^{-3}),\\
	A & = & A_0(u, \phi) + (B_0^2 e^{-2E_0-\varphi}\d_u\d_\phi\varphi +
	B_0 K_4 e^{-2E_0} + A_0^2B_0)r'^{-1} + O(r'^{-2}),\\
	E & = & E_0(u, \phi) + (B_0^2 e^{-E_0} \d_u\varphi - 2 A_0B_0) r'^{-1}
	+ O(r'^{-2}),\\
	u & = & u_0(u, \phi) - B_0^2 e^{-E_0}r'^{-1} + O(r'^{-2}),\\
	\phi & = & \phi_0(u, \phi) - 2 B_0 e^{-\varphi-E_0} r'^{-1} + O(r'^{-2}).
\end{eqnarray*}
At this stage, we have fixed the radial dependence of all the unknowns
and are left with six functions $A_0, B_0, E_0, u_0, \phi_0$, $\chi_0$
of $u'$ and $\phi'$.

We now have to require $l'_\mu = \delta^{u'}_\mu$. After having imposed
$l'=\ddl{}{r'}$, one has in particular that $l'_{r'}=0$. This follows
from the combination of $l'_{r'}={\Lambda^2}_b{e^b}_\nu\ddl{x^\nu}{r'}$ and
$\ddl{x^\nu}{r'}={\Lambda_1}^c{e_c}^\nu$.  For
the remaining components of $l'_\mu$ it is enough to verify that
$l'_{u'}=1+o({r'}^0)$, $l'_{\phi}=o({r'}^0)$ since the equation of motion
$de'^{a}+{\Gamma'^{a}}_be'^{b}=0$ for $a=2$ implies $\d_{r'}l_{\nu'}-\d_{\nu'}
l'_{\mu'}{e_1}^{\mu'}+\Gamma'_{1b1}{e'^{b}}_{\nu'}-\Gamma'_{11\nu'}=0$. This
reduces to $\d_{r'} l'_{\nu'}=\d_{\nu'} l'_{r'}$, and thus to $\d_{r'}
l'_{u'}=0=\d_{r'} l'_{\phi}$. Extracting the leading order from  
\begin{equation}
	e'^2_\mu = (-c^2 W + d^2) \frac{\d u}{\d x'^\mu} + c^2  \frac{\d r}{\d
	x'^\mu} +
	(-c^2 U +cd)\frac{1}{T}\frac{\d \phi}{\d x'^\mu},
\end{equation}
we get
\begin{equation}
	\label{eq:3DgaugesysI}
	e^{-E_0} =
	\frac{\d
	u_0}{\d u'} - B_0 e^\varphi \frac{\d\phi_0}{\d u'},\quad
	0  = \frac{\d
	u_0}{\d \phi'} - B_0 e^\varphi \frac{\d\phi_0}{\d \phi'}.
\end{equation}
We still have to impose three conditions: $V' = O(r'^{-1})$, $\sigma'
= -\frac{1}{2r'} + O(r'^{-3})$, $\tau' = O(r'^{-2})$. The first one is
a condition on the triad and can be imposed by requiring $e'^3_u =
O(1)$. More generally, we have
\begin{equation}
	e'^3_\mu = (-2ac W + 2bd) \frac{\d u}{\d x'^\mu} + 2ac  \frac{\d r}{\d
	x'^\mu} +
	(-2ac U + ad +bc)\frac{1}{T}\frac{\d \phi}{\d x'^\mu},
\end{equation}
and, requiring the new cotriads to have the same form in the new
coordinate system than they had in the old one, the leading terms of $e'^3_u$
and $e'^3_\phi$ yield
\begin{eqnarray}
	\label{eq:3DgaugesysII}
	0 = 
	 e^\varphi \frac{\d\phi_0}{\d u'},\quad
	e^{-E_0+\varphi'}  = e^\varphi \frac{\d\phi_0}{\d \phi'}.
\end{eqnarray}
Note in particular that our choice of parametrization for the Lorentz
rotations leads to $\ddl{\phi_0}{\phi'}>0$. The first equation is
equivalent to $V'=O(r'^{-1})$ while the second one gives the
transformation law of $\varphi$. Combining \eqref{eq:3DgaugesysII}
with \eqref{eq:3DgaugesysI}, we obtain
\begin{equation}
	\label{eq:bms3multstuff}
	\frac{\d\phi_0}{\d u'} = 0, \quad \frac{\d u_0}{\d u'} = e^{-E_0},
	\quad B_0 = e^{-\varphi} (\frac{\d \phi_0}{\d \phi'})^{-1}
	\frac{\d u_0}{\d \phi'}, \quad e^{\varphi'} = e^{E_0+\varphi}
        \frac{\d\phi_0}{\d \phi'}. 
\end{equation}
To implement the last two conditions, we will use the
transformation law of $\widecheck\omega_3$ given in the last equation of
\eqref{eq:434da}. Imposing $(\widecheck\omega_3')_{11} = O(r'^{-2})$ and
$(\widecheck\omega_3')_{21} = \frac{1}{2r'} + O(r'^{-3})$, we get
\begin{equation}
\begin{split}
	 & \frac{1}{2r'}\left( B_0 e^{-E_0} \d_u \varphi - A_0\right) +
	\frac{1}{2} \delta'E + B \delta'A = O(r'^{-2}),\\
&  \frac{1}{r'^2} \left(-\half \chi_0 + 2 A_0B_0 - B_0^2
e^{-E_0}\gamma\right)  \\ & \qquad -B^2 \delta'A - B(1+AB) \delta'E +
	\delta' B = 	O(r'^{-3}).
\end{split}
\end{equation}
From the general solution, we have $\delta'
= U'\d'_{r} + T'\d'_{\phi}$ where $U' = O(1)$ and $T' =
e^{-\varphi'}r'^{-1}+O(r'^{-2})$. Inserting this into the two equations we
can extract the value of $A_0$ and $\chi_0$,
\begin{equation}
	A_0 = e^{-\varphi'} \frac{\d E_0}{\d\phi'}+ B_0 e^{-E_0}
	\frac{\d\varphi}{\d u},\quad
	\chi_0 = 2 A_0B_0+ 2 e^{-\varphi'}\frac{\d B_0}{\d \phi'}.
\end{equation}

The asymptotic symmetry group is thus parametrised by three functions
$u_0(u',\phi')$, $\phi_0(u',\phi')$ and $E_0(u',\phi')$ satisfying the
constraints
\begin{equation}
	\label{eq:changeBMSfinal}
	\frac{\d\phi_0}{\d u'} = 0, \quad \frac{\d u_0}{\d u'} = e^{-E_0}.
\end{equation}
Note that, when taking these into account, the Jacobian matrices for
the change of coordinates at infinity are
\begin{equation}
  \label{eq:47}
  \begin{pmatrix} \ddl{u_0}{u'}=e^{-E_0} & \ddl{u_0}{\phi'}\\
\ddl{\phi_0}{u'}=0 & \ddl{\phi_0}{\phi'}=e^{\varphi'-\varphi-E_0}
\end{pmatrix},\ \begin{pmatrix} \ddl{u'_0}{u}=e^{E_0} &
  \ddl{u'_0}{\phi}=-e^{2E_0-\varphi'+\varphi}\ddl{u_0}{\phi'}\\
\ddl{\phi'_0}{u}=0 & \ddl{\phi'_0}{\phi}=e^{E_0-\varphi'+\varphi}
\end{pmatrix}
\end{equation}

For notational simplicity, we will drop the subscript $0$ on the
functions determining the change of coordinates at infinity and on the
Weyl parameter in the next two section.

\subsection{Combined BMS3 and Weyl group}
\label{sec:geometry-scri}

From equation \eqref{eq:changeBMSfinal}, it follows that
$E(u',\phi')$ is determined by the function $u(u',\phi')$ and,
conversely, that the knowledge of such a function $E$ allows one to
recover the complete change of coordinates, up to an arbitrary
function $\hat u'(\phi')$,
\begin{equation}
	u(u', \phi') = \int^{u'}_{\hat u'} dv'\, e^{-E}. 
\label{eq:u}
\end{equation}
Note that the point with coordinates $(\hat u'(\phi'),\phi')$ in the new
coordinate system is described by
$(0,\phi)$ in the original coordinate system. When considering the inverse
transformation, we can write, 
\begin{equation}
	u'(u, \phi) = \int^u_{\hat u} dv \, e^{E},
\end{equation}
where $E_0$ is now considered as a function of the original coordinate system
through $E_0(u'(u,\phi),\phi'(u,\phi))$ and the point with
coordinates $(\hat u(\phi), \phi)$ is described by 
$(0, \phi')$ in the new coordinate system. 

Equation \eqref{eq:bms3multstuff} is equivalent to the transformation
law of field $\varphi$,  
\begin{equation}
	\label{eq:bms3transvarphi}
	\varphi'(u',\phi')=\varphi(u,\phi)+E(u',\phi')+\ln{\ddl{\phi}{\phi'}}.
\end{equation}
This can be used to trade $\hat u(\phi)$ for
\begin{equation}
\label{eq:beta}
\beta(\phi) = \int^0_{\hat u} dv\, e^{-\varphi(v,\phi)},
\end{equation}
which can be inverted since the integrand is positive.

The combined BMS$_3$ and Weyl group can be parametrized by
\begin{equation}
(\phi'_0(\phi),\beta(\phi),E(u', \phi'))\label{eq:32}.
\end{equation}
Note that 
the transformation law of 
\begin{equation}
  \label{eq:106}
   \quad  \widetilde
u(u,\phi)=\int^{u}_0 dv\ e^{-\varphi(v,\phi)}, 
\end{equation}
is 
\begin{equation}
  \label{eq:57}
  \widetilde u'(u',\phi')=\ddl{\phi'}{\phi}\big[\widetilde u(u,\phi)+\beta(\phi)\big]. 
\end{equation}
In particular, if
\begin{equation*}
  \beta(\phi),E(u',\phi'),\quad
  \beta^{s}(\phi'),E^{s}(u'',\phi''),\quad \beta^{c}(\phi),E^{c}(u'',\phi''),
\end{equation*}
are associated to
a first, a second successive and  their combined transformation respectively,
equations \eqref{eq:bms3transvarphi} and \eqref{eq:106} imply that 
\begin{equation}
  \label{eq:76}
  \begin{split}
    &
    \beta^{c}(\phi)=\ddl{\phi}{\phi'}\beta^{s}(\phi')+\beta(\phi),\\
    &
    E^{c}(u'',\phi'')=E^{s}(u'',\phi'')+E(u',\phi'). 
  \end{split}
\end{equation}
For fixed diffeomorphism on the circle, $\phi'=\phi$, the first of
\eqref{eq:76} describes the abelian subgroup of supertranslations,
while, if in addition one restricts to the subgroup without
supertranslations, i.e., when all $\beta$'s vanish, so do the $\hat
u$'s and $u$ is unchanged. The second of \eqref{eq:76} then describes
the abelian subgroup of Weyl rescalings.

Alternatively, one can define
\begin{equation}
  \label{eq:46}
  \cU(u,\phi)=e^\varphi\widetilde u,\quad 
  \alpha(u,\phi)=e^\varphi\beta, 
\end{equation}
and parametrize the combined BMS$_3$ and Weyl group by 
\begin{equation}
  \label{eq:89}
  (\phi'(\phi),\alpha(u,\phi),E(u', \phi')).
\end{equation}
In this case, equation (\ref{eq:57}) and the first of equation
(\ref{eq:76}) are replaced by 
\begin{equation}
  \label{eq:88}
\begin{split}
  & \cU'(u',\phi')=e^{E(u',\phi')}\big[\cU(u,\phi)+\alpha(u,\phi)\big],\\
& \alpha^c(u,\phi)=e^{-E(u',\phi')}\alpha^s(u',\phi')+\alpha(u,\phi). 
\end{split}
\end{equation}

Note that if $\varphi$ does not depend on $u$ then
$\widetilde\cU=e^{-\varphi} u$, $\beta=-e^{-\varphi}\hat u(\phi)$
whereas $\cU=u$ and $\alpha(\phi)=-\hat u$. If furthermore $\varphi'$
does not depend on $u'$, then neither does $E$ and
$u'(u,\phi)=e^{E(\phi')}(u+\alpha)$. The standard definition of the
BMS$_3$ group is then recovered when the conformal factor is fixed to
be zero, i.e., when $\varphi(u,\phi)=0=\varphi'(u',\phi')$, in which
case it follows from equation \eqref{eq:bms3transvarphi} that the Weyl
transformations are frozen to $e^{E_F}=\ddl{\phi'}{\phi}$.

\subsection{Action on solution space}
\label{sec:bms3action}

Solution space is parametrized by the three functions $\varphi$,
$K_3$, $K_4$ satisfying the evolution equations
\eqref{eq:bms3evolI}. The action of the group on the conformal factor
$\varphi$ has already been computed in the previous section. We can
extract the transformation law of $K_4$ from $(\widecheck \omega'_3)_{12}$
and the one of $K_3$ from the second order of $e'^1_\phi$,
\begin{equation}
\begin{split}
	K'_4& =  e^{-2E} K_4+A_0^2 + 2 e^{-\varphi'}\d_{\phi'}A_0 + 2
	B_0e^{-2E-\varphi} \d_\phi\d_u \varphi,\\
	K'_3 & =  e^{-2E} K_3 + 2 e^{-2E} B_0 K_4 - 2 e^{-\varphi'}
	\d_{\phi'}(e^{-\varphi'}\d_{\phi'}B_0) \\ & \qquad- 2 A_0 e^{-\varphi'}
	\d_{\phi'} B_0 +2  B_0^2 e^{-2E-\varphi} \d_\phi \d_u \varphi.\label{eq:trs}
\end{split}
\end{equation}
By construction, the transformed quantities have to satisfy the
transformed equations, i.e., equations \eqref{eq:bms3evolI} where all
quantities, $K_3,K_4,\varphi$, $u,\phi$, $\partial_u,\partial_\phi$
are primed.

In the particular case where $\varphi=0$, equations
\eqref{eq:bms3evolI} reduce to $\d_u K_{4R}=0$, $\d_u K_{3R}=\d_\phi
K_{4R}$, so that $K_{4R} = (16\pi G) p(\phi)$, $K_{3R}
= (16\pi G) (j(\phi) + u \d_{\phi}p)$. Applying the particular Weyl
transformation $\phi'=\phi$, $u'= \int_0^u dv\, 
e^{E}$, $E(u',\phi')=\varphi'(u',\phi')$, with inverse transformation $u=\int^{u'}_0dv'
e^{-\varphi'(v',\phi')}$, we
obtain from \eqref{eq:trs} that 
\begin{equation}
K'_4=e^{-2\varphi'}[K_{4R}+2\d^2_{\phi'}\varphi'-(\d_{\phi'}\varphi')^2],\
K'_3=e^{-2\varphi'}[K_{3R}+2\d_{\phi'}u K_{4R}-2\d^3_{\phi'}u]. 
\label{eq:114}
\end{equation}
After removing all primes and writing the inverse transformation as
in \eqref{eq:106}, it follows that the general solution to the
dynamical equations \eqref{eq:bms3evolI} for arbitray $\varphi$ is
given by
\begin{equation}
\begin{split}
	K_4 &= (16\pi G)e^{-2\varphi} \Big[ p(\phi) + \frac{1}{16 \pi G}
	(2\d_\phi^2 \varphi -
(\d_\phi\varphi)^2)\Big],\\
	K_3 &= (16\pi G) e^{-2\varphi} \Big[  j(\phi) +  \widetilde u \d_\phi
	p(\phi) + 2 \d_\phi \widetilde u p(\phi) - \frac{1}{8\pi G} \d_\phi^3 \widetilde
u\Big].
\end{split}
\end{equation}

The final parametrisation of the solution space studied in section
\ref{sec:newman-unti-solution} is given by the conformal factor
$\varphi$ and the two functions $p(\phi)$ and $j(\phi)$. Their
transformation laws under the combined BMS$_3$ and Weyl group is given
by 
\begin{equation}
  \label{eq:55a}
\begin{split}
  & p'(\phi')=(\ddl{\phi}{\phi'})^{2}\Big[p(\phi)+\frac{c_2}{24
    \pi}\{\phi',\phi\}\Big],\quad c_1=0,\quad c_2=\frac{3}{G}, \\ 
  & j'(\phi') = (\ddl{\phi}{\phi'})^{2} \Big[ j(\phi) - 2 p(\phi)
  \d_\phi \beta - \d_\phi p \beta +\frac{c_2}{24 \pi}\d_\phi^3
  \beta+\frac{c_1}{24\pi}\{\phi',\phi\}\Big].
  \end{split}
\end{equation}
The central charges have the characteristic values for asymptotically
flat three-dimensional Einstein gravity
\cite{Barnich:2006avcorr}. These quantities are thus Weyl invariant,
which needs to be the case by construction since a Weyl transformation
applied to $K_3,K_4$ amounts to applying the combined Weyl transformation
to $p,j$ with the associated change of $\widetilde u$.  Their
transformations under the BMS$_3$ group agree with those derived by
different methods in \cite{Barnich:2012rz,Barnich:2015uva}.

\section{4d asymptotically flat spacetimes at null infinity}
\label{sec:asympt-flat-spac}

\subsection{Newman-Penrose formalism in 4d}
\label{sec:4d-newman-penrose}

In four dimensions with signature $(+,-,-,-)$, we use 
\begin{equation}
  \label{eq:4}
  \eta_{ab}= \begin{pmatrix} 0 & 1 & 0 & 0 \\
1 & 0 & 0 & 0 \\
0 & 0 & 0 & -1 \\
0 & 0 & -1 & 0
\end{pmatrix}. 
\end{equation}
The different elements of the
null tetrad are denoted by $e_a =(l,n,m,\xbar m)$, with the associated
directional covariant derivatives denoted by
$(D,\Delta,\delta,\xbar\delta)$.  In particular, 
\begin{equation}
g^{\mu\nu}=l^\mu n^\nu+l^\nu n^\mu-m^\mu \xbar m^\nu-m^\nu\xbar m^\mu,\quad \nabla_a= n_a
D+l_a\Delta-m_a\xbar \delta -\xbar m_a \delta \label{eq:126a}. 
\end{equation}

The 24 independent $\Gamma_{abc}$'s are parametrized through 12
complex scalars, 
\begin{equation}
\begin{array}{c|c|c|c}
  \nabla & m^a \nabla l_a &  \half (n^a\nabla l_a-\xbar m^a\nabla m_a)
  & -\xbar m^a \nabla n_a \\
  \hline
  D & \kappa=\Gamma_{311} & 
 \epsilon= \half (\Gamma_{211}-\Gamma_{431})  & \pi = -\Gamma_{421}
  \\
  \hline
  \Delta & \tau=\Gamma_{312} & 
\gamma= \half(\Gamma_{212}-\Gamma_{432}) & \nu = -\Gamma_{422}
  \\
  \hline
  \delta & \sigma=\Gamma_{313} & 
\beta= \half(\Gamma_{213}-\Gamma_{433})& \mu = -\Gamma_{423}
\\
  \hline
  \xbar \delta & \rho=\Gamma_{314} & 
\alpha= \half(\Gamma_{214}-\Gamma_{434}) & \lambda = -\Gamma_{424}
\end{array}
\end{equation}
where the associated complex conjugates are obtained by exchanging the
indices $3$ and $4$. 

In order to describe Lorentz transformations in four dimensions in
terms of a null tetrad, one associates to a real vector $v=v^a e_a$,
with $v^1,v^2\in \mathbb R, v^4=\overline{v^3}\in \mathbb C$, a
$2\times 2$ hermitian matrix $\widehat v=v^{a}\widehat j_{a}$, where the
$\widehat j_{a}$
are chosen as
\begin{equation}
\label{generatorsb}
\widehat j_1 = \begin{pmatrix} 1&0\\ 0& 0 \end{pmatrix},\:
\widehat j_2 = \begin{pmatrix} 0 & 0\\0 &1 \end{pmatrix},\:
\widehat j_3= \begin{pmatrix} 0 & 1\\ 0&0 \end{pmatrix},\:
\widehat j_4= \begin{pmatrix} 0 & 0 \\ 1&0 \end{pmatrix}. 
\end{equation}
In this case
\begin{equation}
\label{invQuad4d}
{\rm det}\ \widehat v=\half \eta_{ab}v^av^b, \, \widehat j^T_b \epsilon \widehat j_a +
\widehat j^T_a\epsilon \widehat j_b =
\eta_{ab}\epsilon,\, \widehat j_b \epsilon \widehat j^T_a +
\widehat j_a\epsilon \widehat j^T_b =
\eta_{ab}\epsilon, \, \widehat j^a(\widehat j_b+\widehat j_b^T)\widehat j^T_a = 0,
\end{equation}
where
\begin{equation}
  \label{eq:21}
  \epsilon= \begin{pmatrix} 0 & 1 \\ -1&0 \end{pmatrix}. 
\end{equation}
For an element $g\in {\rm SL}(2,\mathbb C)$, one considers the
transformation 
\begin{equation}
  g \widehat j_{a} g^{\dagger} v^{a}= \widehat j_a
  {\Lambda^{a}}_{b}v^{b}, \quad \epsilon g^T=  g^{-1}\epsilon.\label{eq:434db}
\end{equation}
More explicitly, if  
\begin{equation}
  \label{eq:741}
  g= \begin{pmatrix}
a & b \\ c & d
\end{pmatrix},
\end{equation}
with $ad-bc=1$ and $a,b,c,d\in \mathbb C$, then
\begin{equation}
\label{Phi4d}
\Lambda^{a}_{\;\;b}=
\begin{pmatrix}
a\xbar a & b\xbar b & a\xbar b & b \xbar a
 \\
c \xbar c & d\xbar d & c\xbar d &
d\xbar c \\
a\xbar c & b\xbar d & a\xbar d &
b\xbar c \\
c\xbar a & d \xbar b & c \xbar b & d\xbar a
\end{pmatrix},\ \Lambda_{a}^{\;\;b}=
\begin{pmatrix}
 d\xbar d & c\xbar c  & -d\xbar c &
 -c\xbar d \\
 b\xbar b  & a\xbar a & -b\xbar a &
-a\xbar b \\
-d\xbar b & -c\xbar a & d\xbar a &
c\xbar b \\
-b\xbar d & -a\xbar c & b\xbar c &
a\xbar d 
\end{pmatrix},
\end{equation}
where the first index is the lign index. 

The standard three classes of rotations \cite{Chandrasekhar:1985kt}
are then given by
\begin{itemize}
\item class I for which $l'=l, m'=m+Al, \xbar m'=\xbar m+\xbar A l,
  n'=n+\xbar A m+ A\xbar m+A\xbar A l$: \\ $a=1=d$,
  $c=0$, $b=-\xbar A$, $A\in \mathbb{C}$, 

\item class II for which $n'=n, m'=m+Bn, \xbar m'=\xbar m+\xbar B n,
  l'=l+\xbar B m+ B\xbar m+B\xbar B n$: \\ $a=1=d$,
  $b=0$, $c=- B$, $B\in \mathbb{C}$, 

\item class III for which $l'=e^{-E_R}l, n'=e^{E_R}
  n, m'=e^{iE_I}m, \xbar
  m'=e^{-iE_I}\xbar m$: \\
$a= e^{-E/2}$,
  $d= e^{E/2} $, $b=0=c$, $E=E_R+iE_I\in \mathbb
  C$.  
\end{itemize}

Finally, the ${\rm SL}(2,\mathbb C)$ group element corresponding to a
combined rotation $II\circ I\circ III$ is given by 
\begin{equation}
  \label{eq:19a}
  g=\begin{pmatrix}
    e^{-E/2}  &  -\xbar A e^{E/2} \\  -B
    e^{-E/2} & (1+\xbar A B)e^{E/2}
  \end{pmatrix}.
\end{equation}
Defining
\begin{equation}
\widehat\omega=-\frac{1}{2}\widehat j_a\epsilon \widehat j_b^T \Gamma^{ab} =
\begin{pmatrix}
-\Gamma_{42} & -\frac{1}{2} (\Gamma_{21} - \Gamma_{43})\\
-\frac{1}{2} (\Gamma_{21} - \Gamma_{43}) & \Gamma_{31}
\end{pmatrix},
\end{equation}
the transformation law of Lorentz connection becomes
\begin{equation}
\widehat\omega' = g \widehat \omega g^T - g \epsilon d g^T.
\end{equation}
More explicitly, for the spin coefficients encoded in 
\begin{equation}
  \label{eq:84}
  \widehat\omega_1=\begin{pmatrix} \pi & -\epsilon\\ -\epsilon &
    \kappa\end{pmatrix},\: \widehat\omega_2=\begin{pmatrix} \nu & -\gamma\\ -\gamma &
    \tau \end{pmatrix},\: \widehat\omega_3=\begin{pmatrix} \mu & -\beta\\ -\beta &
    \sigma\end{pmatrix},\: \widehat\omega_4=\begin{pmatrix} \lambda & -\alpha\\ -\alpha &
    \rho \end{pmatrix},
\end{equation}
one finds
\begin{equation}
  \label{eq:13a}
  \widehat\omega'_a={\Lambda_a}^c g\widehat\omega_c g^T- g\epsilon
  e'_a(g^T). 
\end{equation}

Alternatively, one can use $\widecheck v = \widehat v \epsilon$ in order
to describe real vectors. The
associated basis is
\begin{equation}
\label{generatorsc}
\widecheck j_1 = \begin{pmatrix} 0&1\\ 0& 0 \end{pmatrix},\:
\widecheck j_2 =\begin{pmatrix} 0 & 0\\ -1& 0 \end{pmatrix},\:
\widecheck j_3=  \begin{pmatrix} -1&0\\ 0&0 \end{pmatrix},\:
\widecheck j_4=  \begin{pmatrix} 0&0\\ 0&1 \end{pmatrix},
\end{equation}
so that 
\begin{equation}
\label{invQuad4d1b}
{\rm det}\ \widecheck v=\half \eta_{ab}v^av^b,\, \widecheck j^T_b \epsilon \widecheck j_a +
\widecheck j^T_a\epsilon \widecheck j_b =
\eta_{ab}\epsilon,\, \widecheck j_b \epsilon \widecheck j^T_a +
\widecheck j_a\epsilon \widecheck j^T_b =
\eta_{ab}\epsilon, \, \widecheck j^a(\widecheck j_b+\widecheck j_b^T)\widecheck j^T_a = 0.
\end{equation}
In this case, we have
\begin{gather}
  g \widecheck j_{a} g^{-1} v^{a}= \widecheck j_a
  {\Lambda^{a}}_{b}v^{b},\label{eq:434dc}\quad
\widecheck \omega' = g \widecheck \omega g^{-1} -d g g^{-1},
\quad \widecheck\omega'_a={\Lambda_a}^c g\widecheck\omega_c g^{-1}- e'_a(g)
  g^{-1},
\end{gather}
where
\begin{equation}
  \label{eq:84a}
  \widecheck\omega_1=\begin{pmatrix} \epsilon & \pi\\ -\kappa & -\epsilon
   \end{pmatrix},\: \widecheck\omega_2=\begin{pmatrix} \gamma & \nu\\
     -\tau & -\gamma 
   \end{pmatrix},\: \widecheck\omega_3=\begin{pmatrix} \beta & \mu\\ -
     \sigma & -\beta\end{pmatrix},\:
   \widecheck\omega_4=\begin{pmatrix}\alpha & \lambda\\ -\rho & -\alpha \end{pmatrix}.
\end{equation}

For the Weyl scalars, we follow the conventions of
\cite{Penrose:1984,Penrose:1986}, which differ by a sign from those of
\cite{Chandrasekhar:1985kt} and those of \cite{Newman1962,Newman1962a}
(when taking into account in addition the correction for $\Psi_2$
given in \cite{Newman1963,newman:1980xx}). If $C_{abcd}$ denote the
components of the Weyl tensor and $\Psi_{ABCD}$ the associated Weyl
spinor,
\begin{equation}
  \label{eq:97}
  \begin{split}
    & \Psi_0=C_{1313}\leftrightarrow \Psi_{0000},\quad
    \Psi_1=C_{1213}\leftrightarrow \Psi_{0001},\quad
    \Psi_2=C_{1342}\leftrightarrow \Psi_{0011},\\
    & \Psi_3=C_{1242}\leftrightarrow \Psi_{0111},\quad
    \Psi_4=C_{2424}\leftrightarrow \Psi_{1111}.
  \end{split}
\end{equation}
Their transformations law under Lorentz rotations can be either worked
out directly by using
$C'_{a_1a_2a_3a_4}={\Lambda_{a_1}}^{b_1}\dots{\Lambda_{a_4}}^{b_4}C_{b_1b_2b_3b_4}$
and the symmetries of the Weyl tensor, as done in
\cite{Chandrasekhar:1985kt} for the individual rotations of type
$I,II,III$. A faster way is to use the correspondence with the Weyl
spinor: with our choice of Infeld-van der Waerden symbols in
(\ref{generatorsb}) (cf.~(3.1.50) of \cite{Penrose:1984}),
$e_a=(l,n,m,\xbar m)\leftrightarrow \epsilon_A\xbar \epsilon_{\dot
  A}=(\epsilon_0\xbar \epsilon_0,\epsilon_1\xbar
\epsilon_1,\epsilon_0\xbar \epsilon_1, \epsilon_1\xbar \epsilon_0)$,
$e'_a={\Lambda_a}^be_b$ with ${\Lambda_a}^b$ as in \eqref{Phi4d}
corresponds to $\epsilon'_{A}={g_{A}}^B\epsilon_B$ with
\begin{equation}
  \label{eq:98}
{g_{A}}^B=\begin{pmatrix} -d & c \\ b & -a    
  \end{pmatrix},\quad {g^{A}}_B=\begin{pmatrix} a & b \\ c & d    
  \end{pmatrix}
\end{equation}
When taking into account the complete symmetry of the Weyl spinor, one
gets directly from
$\Psi'_{A_1A_2A_3A_4}={g_{A_1}}^{B_1}\dots{g_{A_4}}^{B_4}\Psi_{B_1B_2B_3B_4}$ 
that
\begin{equation}
\begin{split}
  \Psi'_4 & = a^4 \Psi_4 - 4 a^3 b
  \Psi_3 + 6 a^2b^2 \Psi_2 -4 ab^3 \Psi_1+ b^4 \Psi_0, \\
  \Psi'_3 &= -a^3c \Psi_4 + (a^3d + 3a^2bc) \Psi_3 - 3(ab^2c+a^2bd)
  \Psi_2 \nonumber \\ & \qquad + (b^3c + 3 ab^2d) \Psi_1- b^3d \Psi_0,
  \\
  \Psi'_2& =a^2c^2 \Psi_4 - 2 (abc^2 + a^2cd) \Psi_3 + (b^2c^2 + 4
  abcd + a^2d^2) \Psi_2 \nonumber \\ & \qquad -2 (b^2cd +
  abd^2)\Psi_1+b^2d^2 \Psi_0,
  \\
  \Psi'_1 & = - ac^3 \Psi_4 + (bc^3 + 3 ac^2d) \Psi_3- 3(bc^2d +acd^2)
  \Psi_2 \nonumber \\ & \qquad +(3bcd^2+ad^3)\Psi_1- bd^3\Psi_0,\\
  \Psi'_0 & = c^4 \Psi_4 - 4c^3d\Psi_3 + 6 c^2d^2 \Psi_2 - 4 c d^3
  \Psi_1+d^4 \Psi_0.
      \end{split}
    \end{equation}

\subsection{Newman-Unti solution space}
\label{sec:gauge-fixing-cond}

The gauge fixing conditions at null infinity\footnote{We restrict the
  discussion to $\scrip$.} that are usually assumed correspond to
imposing the six real conditions encoded in
$\kappa=\epsilon=\pi=0$. This is equivalent to requiring
$\Gamma_{ab1}=0$ and can be achieved by a suitable Lorentz
rotation. According to the definition of the Newman-Penrose scalars,
it implies that the whole tetrad is parallely transported along $l$,
$D l=0=D n= Dm= D\xbar m$. In particular, this means that $l$ is the
generator of affinely parametrized null geodesics. One then
requires in addition that $l$ is hypersurface orthonormal and a
gradient, which yields 3 more conditions, $\rho=\xbar \rho$ and
$\tau=\xbar\alpha+\beta$, see, e.g., section 1.9 of
\cite{Chandrasekhar:1985kt}. 

This allows one to choose Bondi coordinates $x^\mu=(u,r,x^A)$,
$\mu=0,\dots,3$, $A=2,3$, $x^A=(\zeta,\xbar\zeta)$ such that the
surfaces $u={\rm cte}$ are null with normal vector $l$,
$l_\mu=\delta_\mu^0$ and that $r$ is the suitably normalized affine
parameter on the null geodesics generated by $l$. The tetrad then
takes the form 
\begin{equation}
l=\ddl{}{r},\quad n=\ddl{}{u}+U\ddl{}{r}+X^A\ddl{}{x^A},\quad
m=\omega\ddl{}{r}+\xi^A\ddl{}{x^A},\label{eq:8}
\end{equation}
which implies that
\begin{equation}
  g^{0\mu}=\delta^\mu_1,\ g^{11}=2(U-\omega\xbar\omega),\
  g^{1A}=X^A-(\xbar\omega\xi^A+\omega\xbar\xi^A),\ 
g^{AB}=-(\xi^A\xbar\xi^B+\xi^B\xbar\xi^A)\label{eq:10}.
\end{equation}
Note furthermore that if $\xi_A=g_{AB}\xi^B$ with $g_{AB}$ the two
dimensional metric inverse to $g^{AB}$, then $\xi^A\bar \xi_A=-1$,
$\xi^A\xi_A=0=\bar\xi^A\bar\xi_A$. 
The associated cotetrad is given by 
\begin{equation}
  \label{eq:17}
\begin{split}
  e^1=-[U+X^A(\omega\xbar\xi_A+\xbar\omega\xi_A)]du+dr +
  (\omega\xbar\xi_A+\xbar\omega\xi_A)dx^A,\\
e^2=du,\quad e^3=X^A\xbar\xi_Adu-\xbar\xi_Adx^A,\quad e^4=X^A\xi_Adu-\xi_Adx^A.
\end{split}
\end{equation}
On a space-like cut of $\scri^+$, we use coordinates
$\zeta,\xbar\zeta$, and the metric 
\begin{equation}
d\xbar s^2=-\xbar\gamma_{AB}
dx^Adx^B=-2(P\xbar P)^{-1}d\zeta d\xbar\zeta, \label{eq:81}
\end{equation}
with $P\xbar P>0$. For the unit sphere, we have
$\zeta=\cot{\frac{\theta}{2}}e^{i\phi}$ in terms of standard spherical
coordinates and 
\begin{equation}
P_{S}(\zeta,\xbar\zeta)=\frac{1}{\sqrt
  2}(1+\zeta\xbar\zeta)\label{eq:96}. 
\end{equation}
The covariant derivative on the $2$ surface is then encoded in the
operator
\begin{equation}
\eth \eta^s= P\xbar P^{-s}\xbar \d(\xbar P^s\eta^s),\qquad \xbar \eth
\eta^s=\xbar P P^{s}\d(P^{-s}\eta^s)\,, 
\label{eq:34}
\end{equation}
where $\eth,\xbar\eth$ raise respectively lower the spin weight by one
unit.  The weights of the various quantities used here are given in
table \ref{t1}. Complex conjugation transforms the spin weight into
its opposite and leaves the conformal weight unchanged.
\begin{table}[h]
\caption{Spin and conformal weights}\label{t1}
\begin{center}
  \begin{tabular}{c|c|c|c|c|c|c|c|c|c|c|c|c} & $\eth$ & $\d_u$ &
    $\gamma^0$ & $\nu^0$ & $\mu^0$ & $\sigma^0$ & $\lambda^0$ &
    $\Psi^0_4 $&  $\Psi^0_3$ & $\Psi^0_2$ & $\Psi^0_1$ & $\Psi^0_0$  \\
    \hline
    $s$ & $1$ & $0$ & $0$ &  $-1$ & $0$ & $2$ &  $-2$  & $-2$  & $-1$ & $0$ & $1$ &  $2$  \\
    \hline
    $w$ & $-1$ & $-1$ & $-1$ &  $-2$ & $-2$ & $- 1$  & $-2$  & $-3$ & $-3$ & $-3$ & $-3$ & $-3$ \\
\end{tabular}
\end{center} 
\end{table}
Note that $P$ is of spin weight $1$ and ``holomorphic'', $\xbar\eth P=0$ and that
\begin{equation} 
[\xbar \eth, \eth]\eta^s=\frac{s}{2}  R\,
  \eta^s\,,\label{eq:35}
\end{equation}
with $R=2P\xbar P\d \xbar \d \ln (P\xbar P)=2\xbar\eth\eth \ln (P\xbar
P) $, $R_S=2$.  We also have 
\begin{equation}
  \label{eq:82}
  [\d_u,\eth]\eta^s=(\d_u\ln P\eth+s\eth\d_u\ln \xbar P)\eta^s. 
\end{equation}

According to \cite{Newman1962,Newman1963,Newman1962a}, once the
conditions $\kappa=\epsilon=\pi=0$ are fixed and coordinates
$u,r,\zeta,\xbar\zeta$ such that $l_\nu=\delta^u_\nu$,
$l^\nu=\delta^\nu_r$ are chosen, which implies in particular also
that $\rho-\xbar\rho=0=\tau-\xbar\alpha-\beta$, the leading part of
the asymptotic behaviour given in (\ref{eq:78}) follows from the
equations of motion, the condition $\Psi_0=\Psi^0_0r^{-5}+O(r^{-6})$
and uniform smoothness, i.e., a standard restriction on the functional
space imposing how the fall-off conditions in $r$ behave with respect
to differentiation. In addition, the choice of a suitable radial
coordinate is used to put to zero the term in $\rho$ of order
$r^{-2}$, while by a choice of coordinates $x^A$, the leading part
$r^2$ of the spatial metric is set to be conformally flat, and the
constant part of $X^A$ to vanish. Finally, the leading order $r^{-1}$
of $\tau$ is set to zero by a suitable null rotation. As will be
explicitly seen below, these conditions guarantee that the asymptotic
symmetry group is the extended BMS group combined with complex
rescalings.

For the explicit form of asymptotic solution space, we will follow
closely \cite{Newman1962a} (see also \cite{Newman1962,Newman1963}),
except that the complex $P$ used here is twice the $P$ used there and
the $\eth$ operator is taken to agree with the definition used in
\cite{Penrose:1986}. Furthermore, $\zeta=x^3+ix^4$ and
$\nabla=2\xbar\d$. More details can be found for instance in the
reviews \cite{newman:1980xx,Penrose:1984,Penrose:1986,stewart:1991}
and also in \cite{Barnich:2011ty}, where a translation to results in
the BMS gauge as used in \cite{Barnich:2010eb} can be found. Note also
that, as compared to
\cite{Barnich:2010eb,Barnich:2011ty,Barnich:2011mi,Barnich:2013axa},
we have changed the signature of the metric in order to agree with the
standard conventions used in the context of the Newman-Penrose
formalism and that $x^4\to -x^4$.

The asymptotic expansion of on-shell spin coefficients, tetrads and
the associated components of the Weyl tensor are given by
\begin{equation}
  \label{eq:78}
\begin{split}
&  \Psi_0= \Psi^0_0 r^{-5}+O(r^{-6}),\\
& \Psi_1 = \Psi^0_1 r^{-4}-\xbar \eth  \Psi^0_0 r^{-5}+O(r^{-6}),\\
& \Psi_2= \Psi^0_2r^{-3}-\xbar \eth  \Psi^0_1 r^{-4}+O(r^{-5}),\\
& \Psi_3= \Psi^0_3r^{-2}-\xbar \eth  \Psi^0_2 r^{-3}+O(r^{-4}),\\
& \Psi_4= \Psi^0_4r^{-1}-\xbar \eth  \Psi^0_3 r^{-2}+O(r^{-3}),
\end{split}
\end{equation}
\begin{equation}
\begin{split}
& \rho= -r^{-1}-\sigma^0\xbar\sigma^0 r^{-3}+O(r^{-5}),\\
& \sigma= \sigma^0
r^{-2}+(\xbar\sigma^0\sigma^0\sigma^0-\half\Psi^0_0)r^{-4}+O(r^{-5}),\\
& \alpha= \alpha^0 r^{-1}+ \xbar\sigma^0\xbar\alpha^0
r^{-2}+\sigma^0\xbar\sigma^0\alpha^0 r^{-3} +O(r^{-4}),\\
& \beta= -\xbar\alpha^0
r^{-1}-\sigma^0\alpha^0r^{-2}-(\sigma^0\xbar\sigma^0\xbar\alpha^0+\half
\Psi^0_1)r^{-3}+O(r^{-4}),\\
& \tau= -\half \Psi^0_1 r^{-3}+\frac{1}{3}(\half
\sigma^0\xbar\Psi^0_1+\xbar\eth\Psi^0_0)r^{-4}+O(r^{-5}),\\
& \lambda= \lambda^0r^{-1}-\mu^0\xbar\sigma^0r^{-2}
+(\sigma^0\xbar\sigma^0\lambda^0+\half\xbar\sigma^0
\Psi^0_2)r^{-3} +O(r^{-4}),\\
& \mu=\mu^0
r^{-1}-(\sigma^0\lambda^0+\Psi^0_2)r^{-2}
+(\sigma^0\xbar\sigma^0\mu^0+\half\xbar
\eth\Psi^0_1)r^{-3}+O(r^{-4}), \\
& \gamma= \gamma^0 -\half
\Psi^0_2r^{-2}+\frac{1}{6}(2\xbar\eth\Psi^0_1+\alpha^0\Psi^0_1-
\xbar\alpha^0\xbar\Psi^0_1)r^{-3}+O(r^{-4}),\\
& \nu= \nu^0-\Psi^0_3r^{-1}+\half\xbar\eth\Psi^0_2r^{-2}+O(r^{-3}),
\end{split} 
\end{equation}
\begin{equation}
\begin{split}
& U= -(\gamma^0+\xbar\gamma^0)r+\mu^0-\half(\Psi^0_2+\xbar\Psi^0_2)r^{-1}
+\frac{1}{6}(\xbar\eth\Psi^0_1+\eth\xbar\Psi^0_1)r^{-2}+O(r^{-3}),\\ 
& X^\zeta=\xbar{X^{\xbar\zeta}}= \frac{1}{6}\xbar P\Psi^0_1r^{-3}+O(r^{-4}),\\
& \xi^\zeta=\xbar{\xbar\xi^{\xbar\zeta}}= -\xbar P\sigma^0
r^{-2}+O(r^{-4}),\\
& \xi^{\xbar\zeta}=\xbar{\xbar\xi^{\zeta}}= P(r^{-1}+\sigma^0\xbar\sigma^0
r^{-3})+O(r^{-4}), \\
& \omega= \xbar\eth\sigma^0
r^{-1}-(\sigma^0\eth\xbar\sigma^0+\half\Psi^0_1)r^{-2} +O(r^{-3}),
\end{split}
\end{equation}
where
\begin{equation}
  \label{eq:108}
\begin{split}
  & \alpha^0=\half \xbar P \d \ln P,\quad \gamma^0= -\half \d_u\ln
    \xbar P,\\ 
& \mu_0= -\half \xbar\eth\eth \ln (P\xbar
    P)=-\half P\xbar P \xbar\d\d \ln (P\xbar
    P)=-\frac{R}{4},
  \end{split}
\end{equation}
\begin{equation}
  \label{eq:79}
  \begin{split}
    & \lambda^0=(\d_u +3\gamma^0-\xbar\gamma^0){\xbar\sigma}^0,\quad
    \nu^0=\xbar \eth
    (\gamma^0+\xbar\gamma^0), \\
    & \Psi^0_2-\xbar\Psi^0_2=
    \xbar\eth^2\sigma^0-\eth^2\xbar\sigma^0
    +\xbar\sigma^0\xbar\lambda^0
    -\sigma^0\lambda^0,\\
    & \Psi^0_3=-\eth\lambda^0+\xbar\eth
    \mu^0,\quad \Psi^0_4= \xbar \eth\nu^0-(\d_u+4\gamma^0)\lambda^0,
  \end{split}
\end{equation}
and
\begin{equation}
  \label{eq:80}
  \begin{split}
    & (\d_u+\gamma^0+5\xbar\gamma^0)\Psi^0_0= \eth\Psi^0_1+3\sigma^0\Psi^0_2,\\
    & (\d_u+2\gamma^0+4\xbar\gamma^0)\Psi^0_1= \eth\Psi^0_2+2\sigma^0\Psi^0_3,\\
    & (\d_u+3\gamma^0+3\xbar\gamma^0)\Psi^0_2= \eth\Psi^0_3+\sigma^0\Psi^0_4.
  \end{split}
\end{equation}
In this approach to the characteristic initial value problem, freely
specifiable initial data at fixed $u_0$ is given by
$\Psi_0(u_0,r,\zeta,\xbar\zeta)$ in the bulk (with the assumed
asymptotics given above) and by $(\Psi^0_2+\xbar\Psi^0_2)(u_0,\zeta,
\xbar\zeta)$, $\Psi^0_1(u_0,\zeta,\xbar\zeta)$ at $\scri^+$.  The news
$\sigma^0(u,\zeta,\xbar\zeta)$ is free data at $\scri^+$ for all $u$
and determines, together with the other initial data at $\scri^+$,
the would-be conserved BMS currents.

As in \cite{Held1970} (see also \cite{Eastwood:1982aa}), for a field
$\eta^{s,w}$ of spin and conformal weights $(s,w)$, one can
associate a field $\widetilde\eta^{h,\bar h}$ of conformal dimensions
$(h,\bar h)$ through
\begin{equation}
\eta^{s,w}=\xbar P^h P^{\bar h}\widetilde\eta^{h,\bar h},\quad (h,\bar
h)=\big(-\half(s+w),\half(s-w)\big).
\label{eq:107}
\end{equation}
The conformal dimensions of the various quantities used here are
given in table \ref{t2}.
\begin{table}[h]
\caption{Conformal dimensions}\label{t2}
\begin{center}
  \begin{tabular}{c|c|c|c|c|c|c|c|c|c|c|c|c} & $\xbar\d$ & $(P\xbar P)^{-\half}\d_{u}$ &
    $\widetilde\gamma^0$ & $\widetilde\nu^0$ & $\widetilde\mu^0$ &
    $\widetilde\sigma^0$ & $\widetilde\lambda^0$ &
    $\widetilde\Psi^0_4 $&  $\widetilde\Psi^0_3$ &
    $\widetilde\Psi^0_2$ & $\widetilde\Psi^0_1$ & $\widetilde\Psi^0_0$
    \\ 
    \hline $h$ & $0$ & $1/2$ & $1/2$ & ${3}/{2}$ & $1$ & $-1/2$ & $2$ &
    ${5}/{2}$ & $2$ &
    ${3}/{2}$ & $1$ &  $1/2$  \\
    \hline $\xbar h$ & $1$ & $1/2$ & $1/2$ & $1/2$ & $1$ & ${3}/{2}$ & $0$
    &
    $1/2$ & $1$ & ${3}/{2}$ & $2$ & ${5}/{2}$ \\
\end{tabular}
\end{center} 
\end{table}

When expressed in these quantities, \eqref{eq:79} and \eqref{eq:80} become
\begin{equation}
  \label{eq:79a}
  \begin{split}
    & \widetilde\gamma^0=-\half (P\xbar P)^{-\half}\d_u\ln
    \xbar P, \quad
    \widetilde\nu^0=(\d+\half\d\ln (P\xbar P))
    (\widetilde\gamma^0+\xbar{\widetilde\gamma}^0),\\
    & \widetilde\mu^0=-\half\d\xbar\d\ln{(P\xbar P)},\quad
    \widetilde\lambda^0=(P\xbar P)^{-\half} \d_{ u}
    {\xbar{\widetilde\sigma}}^0,\\
    & \widetilde\Psi^0_2-\xbar{\widetilde\Psi}^0_2=
    (\d^2-\half[\d^2\ln (P\xbar P)+\half(\d\ln (P\xbar
    P))^2])\widetilde\sigma^0\\& \hspace{1.5cm} 
-(\xbar\d^2-\half[\xbar\d^2\ln (P\xbar P)+\half(\xbar\d\ln (P\xbar
    P))^2])\xbar{\widetilde\sigma}^0+\xbar{\widetilde\lambda}^0\xbar{\widetilde\sigma}^0
-\widetilde\lambda^0\widetilde\sigma^0
    ,\\
    & \widetilde\Psi^0_3=-\xbar\d\widetilde\lambda^0+(\d +\d\ln(P\xbar P))\tilde\mu^0,\quad
    \widetilde\Psi^0_4= (\d+\frac{3}{2}\d\ln(P\xbar
    P))\widetilde\nu^0-(P\xbar P)^{-\half}\d_{u}\widetilde\lambda^0,
  \end{split}
\end{equation}
and
\begin{equation}
  \label{eq:80a}
  \begin{split}
    & (P\xbar P)^{-\half}\d_{u}\widetilde\Psi^0_0=
    (\xbar\d+2\xbar\d\ln(P\xbar P))\widetilde\Psi^0_1
    +3\widetilde\sigma^0\widetilde\Psi^0_2,\\
    & (P\xbar P)^{-\half}\d_{u} \widetilde\Psi^0_1=
    (\xbar\d+\frac{3}{2}\xbar\d\ln(P\xbar P))\widetilde\Psi^0_2
    +2\widetilde\sigma^0\widetilde\Psi^0_3,\\
    & (P\xbar P)^{-\half}\d_{u}\widetilde\Psi^0_2=
    (\xbar\d+\xbar\d\ln(P\xbar P))\widetilde\Psi^0_3+\widetilde\sigma^0\widetilde\Psi^0_4,
  \end{split}
\end{equation} 
Below, during the construction of the solution to the evolution
equations \eqref{eq:79} and \eqref{eq:80}, we will construct improved
fields of this type that take due care both of the additional
$u$-dependence and of the inhomogeneous parts of the transformation
laws. 

\subsection{Residual gauge symmetries}
\label{sec:resid-gauge-freed}

The residual gauge symmetries are the combined Lorentz transformations
and coordinate changes that leave on-shell spin coefficients and
tetrads invariant. Since these transformations map solutions to
solutions, once the conditions that determine asymptotic solution
space are preserved, no further restrictions can arise. The change of
coordinates is of the form
\begin{equation}
  \label{eq:36}
  u=u(u',r',x'^A),\quad r=r(u',r',x'^A),\quad x^A=x^A(u',r',x'^A), 
\end{equation}
and the unknowns are $A,\xbar A, B,\xbar B,E,\xbar E,u,r,x^A$ as
functions of $u',r',x'^A$.

Using the $a=1$ component of the transformation law 
\begin{equation}
e^{\prime
  \mu}_a\ddl{x^\nu}{x^{\prime\mu}}={\Lambda_a}^b{e_b}^\nu,\label{eq:6}
\end{equation}
it follows
that imposing $l'^\mu=\delta^{\mu}_{r'}$ is equivalent to replacing
the LHS by $\ddl{x^\nu}{r'}$. This gives $\ddl{x^\nu}{r'}=d\xbar d
l^\nu+c\xbar c n^\nu-(d\xbar c m^\nu+{\rm c.c.})$, or more explicitly, 
\begin{equation}
  \label{eq:37}
  \begin{split}
    & \ddl{u}{r'} = B\xbar B e^{-E_R},\\
    & \ddl{x^A}{r'} = B\xbar B e^{-E_R}X^A+ \left[\xbar B(1+\xbar
      AB)e^{iE_I}\xi^A +
      c.c.\right],\\
    & \ddl{r}{r'} = |1+A\xbar B|^2e^{E_R} +B\xbar B e^{-E_R} U + \left[\xbar
      B(1+\xbar AB)e^{iE_I}\omega + c.c.\right].
  \end{split}
\end{equation}
In order to implement the gauge fixing conditions in the new
coordinate system, or equivalently $\widecheck\omega'_1=0$, we rewrite
the last of (\ref{eq:434dc}) as
\begin{equation}
  g^{-1}e'_a(g)={\Lambda_a}^b \widecheck\omega_b-g^{-1}\widecheck\omega'_ag, 
\end{equation}
and require, for $a=1$,
\begin{equation}
  \label{eq:7}
  g^{-1}\ddl{g}{r'}={\Lambda_1}^b \widecheck\omega_b. 
\end{equation}
More explicitly, the conditions on the rotation parameters are
\begin{equation*}
  dD'a-bD'c={\Lambda_1}^b (\widecheck\omega_b)_{11},\
  dD'b-bD'd={\Lambda_1}^b (\widecheck\omega_b)_{12},\
  aD'c-cD'a={\Lambda_1}^b(\widecheck\omega_b)_{21}, 
\end{equation*}
where
\begin{equation*}
  {\Lambda_1}^b (\widecheck\omega_b)_{11}=c\xbar c\gamma-d\xbar
  c\beta-c\xbar d\alpha= B\xbar B e^{-E_R} \gamma + \xbar B (1+\xbar A B)
  e^{iE_I}\beta + B (1+A\xbar B)e^{-iE_I}
  \alpha,
\end{equation*}
\begin{equation*}
 {\Lambda_1}^b (\widecheck\omega_b)_{12}=c\xbar c\nu-d\xbar
  c\mu-c\xbar d\lambda=B \xbar B e^{-E_R} \nu + \xbar B (1+\xbar A B)
  e^{iE_I}\mu + B (1+A\xbar
  B)e^{-iE_I} \lambda,
\end{equation*}
\begin{equation*}
   {\Lambda_1}^b(\widecheck\omega_b)_{21}=-c\xbar c\tau+d\xbar
  c\sigma+c\xbar d\rho=-B \xbar B e^{-E_R} \tau - \xbar B (1+\xbar A B)
  e^{iE_I}\sigma - B (1+A\xbar B)e^{-iE_I} \rho.
\end{equation*}

Note that the additional equation involving ${\Lambda_1}^b
(\widecheck\omega_b)_{22}=-{\Lambda_1}^b (\widecheck\omega_b)_{11}$ on
the RHS follows from the first equation when using
$ad-bc=1$.  
When suitably combining these equations, one finds
\begin{equation}
\begin{split}
  & \ddl{B}{r'} = -e^{E} {\Lambda_1}^b(\widecheck\omega_b)_{21},\\
  & \ddl{\xbar A}{r'} = -\xbar A^2
  e^{E}{\Lambda_1}^b(\widecheck\omega_b)_{21} - e^{-E}
  {\Lambda_1}^b(\widecheck\omega_b)_{12},\\
  & \ddl{E}{r'}= 2 \xbar A
  e^{E}{\Lambda_1}^b(\widecheck\omega_b)_{21} -2{\Lambda_1}^b
  (\widecheck\omega_b)_{11} .\label{eq:37a}
\end{split}
\end{equation}
The set of equations \eqref{eq:37} and
\eqref{eq:37a} forms a system of differential equations for
the radial dependence of the unknown functions. In order to solve it
asymptotically, we assume that the functions have the following
asymptotic behavior,
\begin{equation}
	r= O(r'), \quad   A, E, u, \phi = O(1), \quad B = O(r'^{-1}).
\end{equation}

We can now trade the unknown $r$ in the last of \eqref{eq:37} for
$\chi = re^{-E_R}$ satisfying
\begin{multline}
  \ddl{\chi}{r'} = |1+A\xbar B|^2+B\xbar B e^{-2E_R}U + \big[\xbar
  B(1+\xbar A 
  B)e^{-\bar E}\omega +\\+
  \chi{\Lambda_1}^b(\widecheck\omega_b)_{11}-\chi\xbar A 
  e^{E}{\Lambda_1}^b(\widecheck\omega_b)_{21})+c.c.\big]=1 +
  O(r'^{-2}). 
\end{multline}
Note that the vanishing of the $O(r'^{-1})$ terms follows from non
trivial cancellations. Except for the equation for $r$, which we have
just discussed, the RHS of \eqref{eq:37} and \eqref{eq:37a} are all
$O(r'^{-2})$. We thus have 
\begin{equation}
\begin{split}
  & A = A_0 + O(r'^{-1}), \quad r'B = B_0 +
  O(r'^{-1}), \quad E = E_0 + O(r'^{-1}),\\
  & u = u_0 + O(r'^{-1}),\quad x^A = x^A_0 +
  O(r'^{-1}), \\ & r = e^{E_{R0}}r' +
  r_1+O(r'^{-1})\iff \chi=r'+\chi_1+(r'^{-1}), \label{asy4d}
\end{split}
\end{equation}
where $A_0,B_0,E_0,u_0,x^A_0,r_1,\chi_1$ are all functions of
$u',x'^A$. 
These fall-offs allow us to write
\begin{equation*}
\begin{split}
&g^{-1}\ddl{g}{r'}={\Lambda_1}^b \widecheck\omega_b\iff
  g(u',r',x'^A)=e^{-\int^\infty_{r'}d\tilde
    r({\Lambda_1}^bg\widecheck\omega_b)(u',\tilde
    r,x'^A)}g_0(u',x'^A),
\\
&\ddl{x^\mu}{r'}={\Lambda_1}^b{e_b}^\nu \iff
  x^\mu(u',r',x'^A)=x^\mu(u',\infty,x'^A)-\int^\infty_{r'}d\tilde
  r({\Lambda_1}^b{e_b}^\mu)(u',\tilde r,x'^A),
\end{split}
\end{equation*}
for $\mu=u,A$, and where $g_0(u',x'^A)=g(u',\infty,x'^A)$,
$x^\mu_0=x^\mu(u',\infty,x'^A)$. Even though it will not be explicitly
needed in the sequel, equations \eqref{eq:37} and \eqref{eq:37a} can
be used to work out the next to leading orders,
\begin{equation*}
\begin{split}
  & u = u_0 - B_0 \xbar B_0 e^{-E_{R0}} r'^{-1} + O(r'^{-2}),\\
  & \zeta = \zeta_0 - B_0 e^{-E_0}\xbar P r'^{-1} +
  O(r'^{-2}),\\
  & B = B_0 r'^{-1} + O(r'^{-2}),\\
  & A = A_0 + \big[e^{-2E_{R0}+iE_{I0}}B_0\xbar
  B_0\xbar\nu^0
  +e^{-2E_{R0}}B_0\xbar \mu^0+e^{-2\bar E_0}\xbar
  B_0\xbar\lambda^0+A_0^2\xbar B_0]r'^{-1} +O(r'^{-2}),\\ 
  & E = E_0 +2[ B_0\xbar B_0 e^{-E_{R0}}\gamma^0
    +B_0e^{-E_0}\alpha^0 -\xbar
    B_0e^{-\bar E_0}\xbar\alpha^0-\xbar
    A_0B_0]r'^{-1}+O(r'^{-2}). 
\end{split}
\end{equation*}
At this stage, the unknowns are fixed up to
$A_0,B_0,E_0,,R_1(\chi_1),u_0,x^A_0$ as functions of
$u',x'^A$. 

We now have to require $l'_\mu=\delta_\mu^{u'}$. After having imposed
$l'=\ddl{}{r'}$, one has in particular that $l'_{r'}=0$. This follows
from $l'_{r'}={\Lambda^2}_b{e^b}_\nu\ddl{x^\nu}{r'}$ on the one hand
and from $\ddl{x^\nu}{r'}={\Lambda_1}^c{e_c}^\nu$ on the other.  For
the remaining components of $l'_\mu$ it is enough to verify that
$l'_{u'}=1+o({r'}^0)$, $l'_{A}=o({r'}^0)$ since solutions are
transformed into solutions under local Lorentz and coordinate
transformations. In particular, $de^{'a}+{\Gamma^{'a}}_be^{'b}=0$, and
for $a=2$, $de'_1 +\Gamma'_{1b}e^{'b}=0$. Contracting with
${e'_1}^\mu$ then implies that $\d_{r'}l_{\nu'}-\d_{\nu'}
l'_{\mu'}{e_1}^{\mu'}+\Gamma'_{1b1}{e^{'b}}_{\nu'}-\Gamma'_{11\nu'}=0$. This
reduces to $\d_{r'} l'_{\nu'}=\d_{\nu'} l'_{r'}$, and thus to $\d_{r'}
l'_{u'}=0=\d_{r'} l'_{A}$. Extracting the leading order from  
\begin{multline}
  {e'^2}_\mu=\big[-c\xbar
  c[U+X^A(\omega\bar\xi_A+\xbar\omega\xi_A)]+d\xbar d+c\xbar d
  X^A\xbar\xi_A+d\xbar cX^A\xi_A\big]\ddl{u}{x'^\mu}+\\+c\xbar c
  \ddl{r}{x'^\mu}+\big[c\xbar
  c(\omega\xbar\xi_A+\xbar\omega\xi_A)-c\xbar d\xbar\xi_A-d\xbar
  c\xi_A\big]\ddl{x^A}{x'^\mu}, 
\end{multline}
we get 
\begin{equation}
  \label{eq:77}
  \begin{split}
    &
    1=e^{E_{R0}}\ddl{u_0}{u'}-B_0P^{-1}e^{\bar E_0}\ddl{\xbar\zeta_0}{u'}-\xbar
    B_0\xbar
    P^{-1}e^{E_0}\ddl{\zeta_0}{u'},\\
    & 0=
    e^{E_{R0}}\ddl{u_0}{\zeta'}-B_0P^{-1}e^{\bar E_0}\ddl{\xbar\zeta_0}{\zeta'}-\xbar
    B_0\xbar
    P^{-1}e^{E_0}\ddl{\zeta_0}{\zeta'},
  \end{split}
\end{equation}
together with the complex conjugate of the last equation. When using
that the change of coordinates needs to be invertible at infinity,
these relations are equivalent to 
\begin{equation}
  \label{eq:83}
  e^{E_{R0}}=\ddl{u'_0}{u},\quad B_0=-e^{-\bar E_0}\eth u'_0,
\end{equation}
together with the complex conjugate of the last relation.

We now need the transformation laws of $\tau$, $\sigma$ and $\rho$,
which are obtained from the matrix components $21$ of the last
equation of \eqref{eq:434dc} for $a=2,3,4$. This gives 
\begin{multline}
  \tau'=a\xbar a(d^2\tau+c^2\nu-2cd \gamma)-b\xbar
  a(d^2\sigma+c^2\mu-2cd\beta)-a\xbar
  b(d^2\rho+c^2\lambda-2cd\alpha)+\\+d\Delta'(c)-c\Delta'(d),\label{eq:tau}
  \end{multline}
\begin{multline}  
\sigma'=-c\xbar a(d^2\tau+c^2\nu-2cd \gamma)+d\xbar
    a(d^2\sigma+c^2\mu-2cd\beta)+c\xbar
    b(d^2\rho+c^2\lambda-2cd\alpha)+\\+d\delta'(c)-c\delta'(d),\label{eq:sigma}
\end{multline}
\begin{multline}
    \rho'=-a\xbar c(d^2\tau+c^2\nu-2cd \gamma)+b\xbar
    c(d^2\sigma+c^2\mu-2cd\beta)+a\xbar
    d(d^2\rho+c^2\lambda-2cd\alpha)+\\+d\xbar\delta'(c)-c\xbar\delta'(d).\label{eq:rho}
\end{multline}
In order to proceed we need the asymptotic behavior of
$\Delta,\delta',\xbar\delta'$. Using $e^{\prime
  \mu}_a={\Lambda_a}^b{e_b}^\nu\ddl{x'^\mu}{x^\nu}$ for $a=2,3$, we
get
\begin{equation}
  \label{eq:38}
  \begin{split}
    & n'^\mu=b\xbar bD(x'^\mu)+a\xbar a\Delta (x'^\mu)-[b\xbar a\delta
    (x'^\mu)+{\rm c.c.}],\\
    & m'^\mu=-d\xbar bD(x'^\mu)-c\xbar a \Delta(x'^\mu)+d\xbar
    a\delta(x'^\mu) +c\xbar b\xbar\delta(x'^\mu).
  \end{split}
\end{equation}
Explicitly, this gives
\begin{equation}
  \label{eq:75}
  \begin{split}
    & n'^{u'}=e^{-E_{R0}}\ddl{u'_0}{u}+O(r'^{-1})=1+O(r'^{-1}), \\
    & n'^{r'}=U'=e^{-E_{R0}}\d_u(\half\ln{P\xbar
      P}-E_{R0})r'+O(r'^0), \\
    & n'^A=X'^A=e^{-E_{R0}}\ddl{x'^A_0}{u}+O(r'^{-1}),\\
    & m'^{u'}=O(r'^{-1}),\\
    & m'^{r'}=\omega'=A_0+B_0e^{-E_{R0}}\d_u(\half\ln (P\xbar
    P)-E_{R0})-e^{-\bar E_0}\eth E_{R0}+O(r'^{-1}),\\
    & m'^A=\xi'^A=(B_0e^{-E_{R0}}\ddl{x'^A_0}{u}
      +e^{-\bar E_0}\eth x'^A_0)r'^{-1} +O(r'^{-2}).
  \end{split}
\end{equation}
On-shell the new tetrads need to have the same form in the new
coordinates than they had in the old. This implies in particular
\begin{equation}
  \label{eq:85}
\begin{split}
  & \ddl{\zeta'_0}{u}=0,\quad P'=e^{-\bar E_0}\eth
    \xbar\zeta'_0, \\
& A_0=e^{-2E_{R0}+iE_{I0}}\d_u(\half\ln (P\xbar
    P)-E_{R0})\eth u'_0+e^{-\bar E_0}\eth E_{R0},
  \end{split}
\end{equation}
together with the complex conjugates of these equations. 
In addition the requirement that the leading part of the metric remains
conformally flat implies 
\begin{equation}
  \label{eq:86}
  \ddl{\zeta'_0}{\xbar\zeta}=0=\ddl{\xbar\zeta'_0}{\zeta}. 
\end{equation}
When used in \eqref{eq:77} this leads to 
\begin{equation}
\ddl{u_0}{u'}=e^{-E_{R0}},\quad
\ddl{u'_0}{\zeta}=-e^{E_{R0}}\ddl{\zeta'_0}{\zeta}\ddl{u_0}{\zeta'}. 
\label{eq:66a} 
\end{equation}

In order to work out the term on the RHS of $\tau'$ in \eqref{eq:tau}
of order $O(r'^{-1})$, one needs in particular ${n'}^{r'}$
above. Requiring this term to vanish gives
\begin{equation}
  \label{eq:5}
  A_0=-\d_{u'}B_0+B_0(i\d_{u'}E_{I0}+\half
  e^{-E_{R0}}\d_u\ln{\frac{P}{\xbar P}}).  
\end{equation}
When using \eqref{eq:83}, this coincides with the second of
\eqref{eq:85}.
 
Requiring that the tems of order $r'^{-2}$ in $\rho'$ in equation
\eqref{eq:rho} vanish yields 
\begin{equation}
  \label{eq:87}
  \chi_1=B_0\xbar B_0e^{-E_{R0}}\d_u\ln (P \xbar P)+A_0\xbar B_0+2\xbar A_0
  B_0+\xbar \eth' B_0-2B_0\xbar\eth'E_{R0}. 
\end{equation}
Finally, to leading order, the transformation law of $\sigma$ in
\eqref{eq:sigma} yields
\begin{equation}
  \label{eq:33}
  \sigma'^0=e^{-E_{R0}+2iE_{I0}}\sigma^0-A_0B_0-\eth'B_0.
\end{equation}

In summary, we see that all the unknowns $A_0,\xbar A_0,B_0,\xbar
B_0,\chi_1(R_1),E_{R0}$ are
determined by the change of coordinates at infinity and by
$E_{I0}$. The Jacobian matrices are given by
\begin{equation}
\begin{split}
  \label{eq:47a}
  & \begin{pmatrix} \ddl{u_0}{u'}=e^{-E_{R0}} & \ddl{u_0}{\zeta'} &
    \ddl{u_0}{\xbar\zeta'}\\
\ddl{\zeta_0}{u'}=0 & \ddl{\zeta_0}{\zeta'}=e^{-\bar E_0}\frac{\xbar
  P}{\xbar P'} &
\ddl{\zeta_0}{\xbar\zeta'}=0 \\
\ddl{\xbar\zeta_0}{u'}=0 & \ddl{\xbar\zeta_0}{\zeta'}=0 &
\ddl{\xbar\zeta_0}{\xbar\zeta'}=e^{-\bar E_0}\frac{P}{P'}
\end{pmatrix}, \\
& \begin{pmatrix} \ddl{u'_0}{u}=e^{E_{R0}} &
  \ddl{u'_0}{\zeta}=-e^{E_{R0}}\ddl{\zeta'_0}{\zeta}\ddl{u_0}{\zeta'} &
  \ddl{u'_0}{\xbar\zeta}=-e^{E_{R0}}\ddl{\xbar\zeta'_0}{\xbar\zeta}\ddl{u_0}{\xbar\zeta'}\\
\ddl{\zeta'_0}{u}=0 & \ddl{\zeta'_0}{\zeta}=e^{E_0}\frac{\xbar
  P'}{\xbar P}& \ddl{\zeta'_0}{\xbar\zeta}=0\\
\ddl{\xbar\zeta'_0}{u}=0 & \ddl{\xbar\zeta'_0}{\zeta}=0 &
\ddl{\xbar\zeta'_0}{\xbar\zeta} =e^{\bar E_0}\frac{P'}{P}
\end{pmatrix}.
\end{split}
\end{equation}
Note that here and in the following, when considered as a function of
$(u,\zeta,\xbar\zeta)$, $E_0$ is explicitly given by
$E_0(u'_0(u,\zeta,\xbar\zeta),\zeta'_0(\zeta),\xbar\zeta'_0(\xbar\zeta))$. Note
also that the right lower corner of \eqref{eq:47a} is equivalent to
the transformation law of $P$,
\begin{equation}
  \label{eq:54a}
  P'(u',\zeta',\xbar\zeta')=P(u,\zeta,\xbar\zeta)
e^{-\bar E_0}\ddl{\xbar\zeta'_0}{\xbar\zeta}, 
\end{equation}
and that preserving $P\xbar P>0$ requires
$\ddl{\zeta'_0}{\zeta}\ddl{\xbar\zeta'_0}{\xbar\zeta}>0$. The
transformation law of the metric in \eqref{eq:81} and of $\eth$ are
given by 
\begin{equation}
\begin{split}
  (d\xbar s^2)'& = e^{2E_{R0}}(d\xbar s^2),\\
  \eth'\eta^s& = 
  e^{-\bar E_0}\Big(\eth-e^{-E_{R0}}\eth u'_0\d_u-s\big[\eth E_0-(e^{-E_{R0}}\eth 
  u'_0)\d_u(E_0-\ln \xbar P)\big]\Big)\eta^s. 
 \label{eq:93}
\end{split}
\end{equation}
In particular, when putting all results together, the subleading term
of the rescaled radial coordinate is given by  
\begin{equation}
  \label{eq:100}
  \chi_1=e^{-3E_{R0}}\eth u'_0\xbar\eth
  u'_0[\d_uE_{R0}+\gamma^0+\xbar\gamma^0]-e^{-2E_{R0}}\xbar\eth\eth u'_0. 
\end{equation}

For notational simplicity, we drop in the next sections the subscript
$0$ on the asymptotic change of coordinates and on the complex Weyl
parameter. 

\subsection{Combined extended BMS4 group with complex
  rescalings} 
\label{sec:penrose-bms_4-group}

From the top left corner of the first matrix of
\eqref{eq:47a}, it follows that $E_R$ is determined by
$u(u',\zeta',\xbar\zeta')$ and, conversely, that the knowledge of such a
function allows one to recover the complete change of coordinates, up
to an arbitrary function $\hat u'(\zeta',\xbar\zeta')$,
\begin{equation}
  \label{eq:117}
  u(u', \zeta',\xbar\zeta') = \int^{u'}_{\hat u'} dv'\, e^{-E_R}. 
\end{equation}
Note that the point with coordinates $(\hat
u'(\zeta',\xbar\zeta'),\zeta',\xbar\zeta')$ corresponds to
$(0,\zeta,\xbar\zeta)$.

After inverting, one can write,  
\begin{equation}
	u'(u, \zeta,\xbar\zeta) = \int^u_{\hat u} dv\, e^{E_R}, 
\label{eq:ua}
\end{equation}
where
$E_R=E_R(u'(v,\zeta,\xbar\zeta),\zeta'(\zeta),\xbar\zeta'(\xbar\zeta))$,
and now the point with coordinates $(\hat
u(\zeta,\xbar\zeta),\zeta,\xbar\zeta)$ is given by
$(0,\zeta',\xbar\zeta')$ in the new coordinate system.

For a field $P$ transforming as in \eqref{eq:54a}, we trade $\hat
u(\zeta,\xbar\zeta)$ for
\begin{equation}
\label{eq:betaa}
\beta(\zeta,\xbar\zeta) = \int^0_{\hat u} dv\,
(P\xbar P)^\half,
\end{equation}
which can be inverted since the integrand is positive. 

The extended BMS$_4$ group combined with complex rescalings can be
parametrized by
\begin{equation}
(\zeta'(\zeta),\xbar\zeta'(\xbar\zeta),\beta(\zeta,\xbar\zeta),
E(u', \zeta',\xbar\zeta'))\label{eq:32a},
\end{equation}
since $\beta(\zeta,\xbar\zeta)$ determines $\hat u(\zeta,\xbar\zeta)$
and one then gets $u'(u,\zeta,\xbar\zeta)$ from
\eqref{eq:ua}. Defining 
\begin{equation}
\widetilde u(u,\zeta,\xbar\zeta)=\int_0^udv (P\xbar P)^\half(v,\zeta,\xbar\zeta),\label{eq:39}
\end{equation}
its transformation law is simply
\begin{equation}
  \label{eq:57a}
  \widetilde u'(u',\zeta',\xbar\zeta')=J^{-\half}
  \big[\widetilde u (u,\zeta,\xbar\zeta)+\beta(\zeta,\xbar\zeta)\big],\quad
  J=\ddl{\zeta}{\zeta'} 
    \ddl{\xbar\zeta}{\xbar\zeta'}. 
\end{equation}
Together with \eqref{eq:54a}, this implies in particular that, if
\begin{equation}
  \label{eq:40}
  \beta(\zeta,\xbar\zeta),\quad E(u',\zeta',\xbar\zeta'),
\end{equation}
and the same quantities with a superscript $s$ and a superscript $c$
are associated to a first, a second successive and the combined
transformation respectively, we have
\begin{equation}
  \label{eq:76a}
  \begin{split}
    & \beta^{c}(\zeta,\xbar\zeta)=J^\half
    \beta^{s}(\zeta',\xbar\zeta')+\beta(\zeta,\xbar\zeta),\\
    &
    E^{c}(u'',\zeta'',\xbar\zeta'')=E^{s}(u'',\zeta'',\xbar\zeta'')
    +E(u',\zeta',\xbar\zeta').
  \end{split}
\end{equation}
Alternatively, one can define
\begin{equation}
  \label{eq:90}
  \begin{split}
    {\cU}(u,\zeta,\xbar\zeta)=(P\xbar P)^{-\half}\widetilde u,\quad
    \alpha(u,\zeta,\xbar\zeta)=(P\xbar P)^{-\half}\beta, 
  \end{split}
\end{equation}
and parametrize the extended BMS$_4$ combined with complex
rescaling through
\begin{equation}
  \label{eq:91}
  (\zeta'(\zeta),\xbar\zeta'(\xbar\zeta),\alpha(u,\zeta,\xbar\zeta),
E(u', \zeta',\xbar\zeta')). 
\end{equation}
Equation (\ref{eq:57a}) and the first of equations
(\ref{eq:76a}) are then replaced by
\begin{equation}
  \label{eq:92}
\begin{split}
  & \cU'(u',\zeta',\xbar\zeta')=e^{E_R(u',\zeta',\bar\zeta')}
  \big[\cU(u,\zeta,\xbar\zeta)+\alpha(u,\zeta,\xbar\zeta)\big], \\
  & \alpha^{c}(u,\zeta,\xbar\zeta)=e^{-E_R(u',\zeta',\bar\zeta')}
  \alpha^{s}(u',\zeta',\xbar\zeta')+\alpha(u,\zeta,\xbar\zeta).
  \end{split} 
\end{equation}

A pure complex rescaling is characterized by 
\begin{equation}
  \label{eq:118}
  \zeta'=\zeta,\quad u'=\int^u_0dv\, e^{E_R},\quad P'=P
  e^{-\bar E}.
\end{equation}

In the case when $P$ does not depend on $u$, $\widetilde u=(P\xbar
P)^\half u$ and $\beta=-(P\xbar P)^\half\hat u$, whereas $\cU=u$ and
$\alpha(\zeta,\xbar\zeta)=-\hat u$. If furthermore $P'$ does not
depend on $u'$, then neither does $E$ and
\begin{equation}
u'(u,\zeta,\xbar\zeta)=e^{E_R(\zeta',\bar\zeta')}(u+\alpha),\quad
e^{-E_R}\eth u'=\eth\alpha+\eth E_R (u+\alpha).\label{eq:103}
\end{equation}
When the conformal factor is fixed,
$P(\zeta,\xbar\zeta)=P_F(\zeta,\xbar\zeta)$ and
$P'(\zeta',\xbar\zeta')=P_F(\zeta',\xbar\zeta')$ for some
fixed function $P_F$ of its arguments, it follows from (\ref{eq:54a})
that complex rescalings are frozen to
\begin{equation}
e^{E}=\frac{\xbar P_F(\zeta,\xbar\zeta)}{\xbar
  P_F(\zeta',\xbar\zeta')}\ddl{\zeta'}{\zeta}\label{eq:104}. 
\end{equation}
In this case, a pure supertranslation is characterized by 
\begin{equation}
  \label{eq:119}
  \zeta'=\zeta,\quad e^{E}=1,\quad
  u'(u,\zeta,\xbar\zeta)=u+\alpha,\quad e^{-E_R}\eth u'=\eth\alpha,
\end{equation}
while a pure superrotation is characterized by 
\begin{equation}
  \label{eq:120}
  \zeta'=\zeta'(\zeta),\ e^{E}=\frac{\xbar
    P_F(\zeta,\xbar\zeta)}{\xbar 
  P_F(\zeta',\xbar\zeta')}\ddl{\zeta'}{\zeta},\
u'=e^{E_R} u,\ e^{-E_R}\eth u'=\eth E_R u.  
\end{equation}

The standard definition of the BMS$_4$ group is then recovered when
(i) standard Lorentz rotations are described through fractional linear
transformations (see e.g.~\cite{Held1970} for details),
\begin{equation}
  \label{eq:94}
  \zeta'=\frac{a\zeta+b}{c\zeta+d},\quad ad-bc=1,\quad a,b,c,d\in
  \mathbb{C}, 
\end{equation}
(ii) the conformal factor is fixed to be that for the unit sphere,
$P_F=P_S$, in which case 
\begin{equation}
  \label{eq:95}
  e^{E_R^S}=\frac{1+\zeta\xbar\zeta}{(a\zeta+b)(\xbar a\xbar \zeta+\xbar
    b)+(c\zeta+d)(\xbar c\xbar\zeta+\xbar d)}, \quad
  e^{iE^S_I}=\frac{\xbar c\xbar\zeta+\xbar d}{c\zeta +d},
\end{equation}
(iii) supertranslation are expanded in spherical harmonics,
$\alpha_S=\sum_{l,m}\alpha^{lm}Y_{lm}(\zeta,\xbar\zeta)$ with
ordinary translations corresponding to the terms with $l=0,1$ and are
explicitly described by 
\begin{equation}
  \label{eq:99}
  \alpha=\frac{A+B\zeta+\xbar
    B\xbar\zeta+C\zeta\xbar\zeta}{1+\zeta\xbar\zeta}, \quad A,C\in
  \mathbb R, B\in \mathbb C. 
\end{equation}

\subsection{Action on solution space}
\label{sec:acti-solut-space}

Putting the results of the previous subsections together, the
transformation law of the data characterizing asymptotic solution
space is contained in
\begin{multline}
  \label{eq:101}
  \sigma'_0=e^{-E_R+2iE_I}\Big[\sigma_0+\eth(e^{-E_R}\eth
  u')-\\-(e^{-E_R}\eth u')(\d_u+\xbar\gamma^0-\gamma^0)(e^{-E_R}
  \eth u')\Big], 
\end{multline}
\begin{multline}
\lambda'^0=e^{-2E}\Big[\lambda^0+(\d_u+3\gamma^0-\xbar\gamma^0)\big[
\xbar\eth(e^{-E_R}\xbar\eth
  u')-\\-(e^{-E_R}\xbar\eth u')(\d_u+\gamma^0-\xbar\gamma^0)(e^{-E_R}
  \xbar\eth u')\big]\Big],
\end{multline}
\begin{equation}
  \label{eq:102}
  \begin{split}
    {\Psi'}_4^0 & =  e^{-3E_R-2iE_I} \Psi_4^0,\\
    {\Psi'}_3^0 & = e^{-3E_R-iE_I} \Big[\Psi_3^0 - e^{-E_R}
      \eth u' \Psi_4^0\Big],\\
    {\Psi'_2}^0 & = e^{-3E_R} \Big[\Psi_2^0 - 2e^{-E_R} \eth u'
      \Psi_3^0 + (e^{-E_R}\eth u')^2 \Psi_4^0\Big],\\
    {\Psi'}_1^0 & = e^{-3E_R+iE_I} \Big[\Psi_1^0 - 3
      e^{-E_R}\eth u' \Psi_2^0 + 3 (e^{-E_R}\eth u')^2\Psi_3^0 -
      (e^{-E_R}\eth u')^3\Psi_4^0\Big],\\
{\Psi'}_0^0 & = e^{-3E_R+2iE_I} \Big[\Psi_0^0 - 4
      e^{-E_R}\eth u' \Psi_1^0 + 6 (e^{-E_R}\eth u')^2\Psi_2^0 -\\ &\hspace{6cm}
      -4(e^{-E_R}\eth u')^3\Psi_3^0+(e^{-E_R}\eth
      u')^4\Psi_4^0\Big].
  \end{split}
\end{equation}
In particular for instance, if $P,P'$ do not depend on $u,u'$, the
transformation law of the asymptotic shear under a pure
supertranslation reduces to
\begin{equation}
\sigma'_0= \sigma_0+\eth^2\alpha,\quad u'=u+\alpha,\quad \zeta'=\zeta, \label{eq:122}
\end{equation}
while the transformation law of the news under a pure superrotation is
\begin{equation}
\lambda'^0=e^{-2E}[\lambda^0+(\xbar\eth^2 E_R-(\xbar\eth E_R)^2],\label{eq:123}
\end{equation}
with $u',E$ given in \eqref{eq:120}. If furthermore, we
work with respect to the Riemann sphere, $P_F=P_R=1$, this reduces to
\begin{equation}
  \lambda'^0=(\ddl{\zeta'}{\zeta})^{-2}[\lambda^0+\half\{\zeta',\zeta\}],
  \quad u'=J^{-\half}u,\quad \zeta'=\zeta'(\zeta), \label{eq:124}
\end{equation}
where the Schwarzian derivative is $\{F,x\}=\d^2_x\ln
(\d_xF)-\frac{1}{2} (\d_x\ln
(\d_xF))^2$. 

Let us now analyze in more details the evolution equations
(\ref{eq:79}). We start with unit scaling factors, $P=\xbar P=1$, so
that in particular the leading part of the metric on a space-like cut
of $\scri^+$ is the one on the Riemann sphere, $d\xbar s^2=-2d\zeta
d\xbar\zeta$. In this case, \eqref{eq:79} and \eqref{eq:80} reduce to
\begin{equation}
  \label{eq:79b}
  \begin{split}
    & \gamma^0_R=0=\nu^0_R=\mu^0_R,\quad \lambda^0_R=
    \dot{\xbar\sigma}_R^0,\\
    & \Psi^0_{2R}-\xbar\Psi^0_{2R}=
    \d^2\sigma^0_R-\xbar\d^2\xbar\sigma_R^0+\dot
    {\sigma}_R^0\xbar\sigma_R^0-\dot
    {\xbar\sigma}_R^0\sigma^0_R,\\
    & \Psi^0_{3R}=-\xbar\d \dot{\xbar\sigma}_R^0,\quad \Psi^0_{4R}=
    -\ddot{\xbar\sigma}_R^0,
  \end{split}
\end{equation}
and
\begin{equation}
  \label{eq:80b}
  \begin{split}
    & \d_{u}\Psi^0_{0R}= \xbar\d\Psi^0_{1R}
    +3\sigma^0_R\Psi^0_{2R},\\
    & \d_{u} \Psi^0_{1R}=\xbar\d\Psi^0_{2R}
    +2\sigma^0_R\Psi^0_{3R},\\
    & \d_{u}\Psi^0_{2R}= \xbar\d\Psi^0_{3R}+\sigma^0_R\Psi^0_{4R},
  \end{split}
\end{equation} 
In a first stage, these equations may be trivially solved in terms of
integration functions
$\widetilde\Psi^0_{aRI}=\widetilde\Psi^0_{aRI}(\zeta,\xbar\zeta)$ for
$a=0,1,2$ as follows:
\begin{equation}
  \label{eq:48a}
  \Psi^0_{2R}=\widetilde\Psi^0_{2RI}+\int^u_0
  dv\,\big[\xbar \d \Psi^0_{3R}+\sigma^0_R\Psi^0_{4R}\big],
\end{equation}
where
\begin{equation}
  \label{eq:115ab}
  \widetilde\Psi^0_{2RI}-\widetilde{\xbar\Psi}^0_{2RI}
=(\d^2\sigma^0_R-\xbar\d^2\xbar\sigma_R^0+\dot
    {\sigma}_R^0\xbar\sigma_R^0-\dot
    {\xbar\sigma}_R^0\sigma^0_R)(0), 
\end{equation}
\begin{equation}
  \Psi^0_{1R}=\widetilde\Psi^0_{1RI}+\int^u_0
  dv\,\big[\xbar \d \Psi^0_{2R}+2\sigma^0_R\Psi^0_{3R}\big], 
\end{equation}
and
\begin{equation}
  \Psi^0_{0R}=\widetilde\Psi^0_{0RI}+\int^u_0
  dv\,\big[\xbar \d \Psi^0_{1R}+3\sigma^0_R\Psi^0_{2R}\big],
\end{equation}
and the expressions for $\Psi^0_{1R}$ in terms of
$\widetilde\Psi^0_{1RI},\widetilde\Psi^0_{2RI},\sigma^0_R,\dot{\xbar\sigma}^0_R$
and for $\Psi^0_{0R}$ in terms of
$\widetilde\Psi^0_{0RI},\widetilde\Psi^0_{1RI},\widetilde\Psi^0_{2RI},
\sigma^0_R,\dot{\xbar\sigma}^0_R$ can be worked out recursively.

For later use, we introduce instead the integration
functions $\Psi^0_{aRI}=\Psi^0_{aRI}(\zeta,\xbar\zeta)$ for
$a=0,1,2$ defined by
\begin{equation}
  \label{eq:48}
  \Psi^0_{2R}=\Psi^0_{2RI}-\xbar\d^2{\xbar\sigma}_R^0
-{\sigma}_R^0\dot{\xbar\sigma}_R^0+\int^u_0
  dv\, \dot\sigma^0_R \dot{\xbar\sigma}^0_R,\quad
  \Psi^0_{2RI}=\xbar\Psi^0_{2RI}, 
\end{equation}
\begin{multline}
  \Psi^0_{1R}=\Psi^0_{1RI}+u\xbar\d\Psi^0_{2RI}-\sigma^0_R\xbar\d\xbar\sigma^0_R
-\half\xbar\d(\sigma^0_R\xbar\sigma^0_R) 
+\\
  +\int^u_0
  dv\,\big[\half(\xbar\d\dot\sigma^0_R\xbar\sigma^0_R
-3\sigma^0_R\xbar\d \dot{\xbar\sigma}^0_R+3\dot\sigma^0_R\xbar\d \xbar\sigma^0_R
-\xbar\d\sigma^0_R \dot{\xbar\sigma}^0_R) -\xbar\d^3\xbar\sigma^0_R\big]
+\\+\xbar\d\int^u_0
  dv\int^v_0
  dw\,\dot\sigma^0_R\dot{\xbar\sigma}^0_R, 
\end{multline}
\begin{multline}
  \Psi^0_{0R}=\Psi^0_{0RI}+u\xbar\d \Psi^0_{1RI}+\half u^2\xbar\d^2
  \Psi^0_{2RI}+\\+\int^u_0 dv\,\big[\sigma^0_R(3
  \Psi^0_{2RI}-3\sigma^0_R\dot{\xbar\sigma}^0_R-4\xbar\d^2{\xbar\sigma}^0_R)
  -2\xbar\d\sigma^0_R\xbar\d{\xbar\sigma}^0_R-\half
  \xbar\d^2\sigma^0_R{\xbar\sigma}^0_R-\half
  \sigma^0_R\xbar\d^2{\xbar\sigma}^0_R\big]+\\ +\int^u_0
  dv\Big[3\sigma^0_R\int^v_0 dw\,
  \dot\sigma^0_R\dot{\xbar\sigma}^0_R+\xbar\d\int^v_0
  dw\,\big[\half(\xbar\d\dot\sigma^0_R\xbar\sigma^0_R
-3\sigma^0_R\xbar\d \dot{\xbar\sigma}^0_R
  +3\dot\sigma^0_R\xbar\d \xbar\sigma^0_R-
  \xbar\d\sigma^0_R \dot{\xbar\sigma}^0_R )-\xbar\d^3\xbar\sigma^0_R\big]\Big]
  +\\+\xbar\d^2\int^u_0 dv\int^v_0 dw\int^w_0
  dx\,\dot\sigma^0_R\dot{\xbar\sigma}^0_R.
\end{multline}
We have
\begin{equation}
  \label{eq:116}
\begin{split}
&  \Psi^0_{2RI}=\widetilde\Psi^0_{2RI}+[\xbar\d^2\xbar\sigma^0_R
+\sigma^0_R\dot{\xbar\sigma}^0_R](0),\\
&
\Psi^0_{1RI}=\widetilde\Psi^0_{1RI}+[\sigma^0_R\xbar\d\xbar\sigma^0_R+\half\xbar\d 
(\sigma^0_R\xbar\sigma^0_R)](0),\\
&\Psi^0_{0RI}=\widetilde\Psi^0_{0RI}. 
\end{split}
\end{equation}

In these terms, the Bondi mass aspect can be chosen to be given by 
\begin{equation}
  \label{eq:125}
  (-4\pi G)  M_R=\Psi^0_{2R}
  +\sigma^0_R\dot{\xbar\sigma}^0_R+\xbar\d^2\xbar\sigma^0_R=\Psi^0_{2RI}+\int^u_0 
  dv\, \dot\sigma^0_R \dot{\xbar\sigma}^0_R.  
\end{equation}
Note that this expression contains the additional term
$\xbar\d^2\xbar\sigma^0_R$ as compared to the more convential choice,
see e.g., equation (4.18) of \cite{Barnich:2013axa} with $f=1$ (up to
a global minus sign due to the change of signature). More details on
this quantity, and more generally on the would-be conserved currents
and their transformation laws, will be given elsewhere
\cite{Barnich:2016}.

In order to generate the general solution to (\ref{eq:79}) and
\eqref{eq:80} for arbitrary scaling factors $P,\xbar P$ from the one
with $P=1=\xbar P$, we apply a pure complex rescaling, without
superrotations nor supertranslations to the solution above, i.e., we
take $e^{E}=\xbar P'^{-1}$, $u'=\int^u_0 dv\, e^{E_R}$,
$\zeta'=\zeta$. In this particular case,
\begin{equation}
  \label{eq:105}
\gamma^0=0,\  e^{-E_R}\eth u'=-\xbar\d' u, \
  \eth=\xbar\d'-(P'\xbar P')^{-\half}\xbar\d'u\d_{u'},\
  \d_u=(P'\xbar P')^{-\half}\d_{u'}.  
\end{equation}
As a consequence, one finds from the transformation laws that the
general solution to (\ref{eq:79}) and
\eqref{eq:80} is given by 
\begin{equation}
  \label{eq:109}
\begin{split}
  & \sigma^0=\xbar
  P^{-\half} P^{\frac{3}{2}}\Big[\sigma^0_R(\widetilde
    u)-\xbar\d^2\widetilde u\Big], \\
& \lambda^0= \xbar P^2\Big[\dot{\xbar\sigma}^0_R(\widetilde 
    u)-\half\big(\d^2\ln (P\xbar P)+ \half(\d\ln (P\xbar
    P))^2\big)\Big], \\ 
&  \Psi^0_4=\xbar
  P^{\frac{5}{2}}P^{\half}\Big[\Psi^0_{4R}(\widetilde u)\Big], \\ & \Psi^0_3=\xbar
  P^{2}P\Big[ \Psi^0_{3R}(\widetilde u)+
\xbar\d\widetilde u\Psi^0_{4R}(\widetilde u)\Big],\\
& \Psi^0_2=\xbar
P^{\frac{3}{2}}P^{\frac{3}{2}}\Big[\Psi^0_{2R}(\widetilde
  u)+2\xbar\d\widetilde u\Psi^0_{3R}(\widetilde u)
+(\xbar\d\widetilde u)^2\Psi^0_{4R}(\widetilde u)\Big],\\
& \Psi^0_1=\xbar PP^2\Big[\Psi^0_{1R}(\widetilde
  u)+3\xbar\d\widetilde u\Psi^0_{2R}(\widetilde u)+3
(\xbar\d\widetilde u)^2\Psi^0_{3R}(\widetilde
    u)+(\xbar\d\widetilde u)^3\Psi^0_{4R}(\widetilde u)\Big],\\
& \Psi^0_0=\xbar P^\half P^{\frac{5}{2}}\Big[\Psi^0_{0R}(\widetilde
  u)+4\xbar\d\widetilde u\Psi^0_{1R}(\widetilde u)+6
(\xbar\d\widetilde u)^2\Psi^0_{2R}(\widetilde u)+ \\ &\hspace{6.5cm} +4(\xbar\d\widetilde u)^3
\Psi^0_{3R}(\widetilde u)+(\xbar\d\widetilde u)^4\Psi^0_{4R}(\widetilde u)\Big],
\end{split}
\end{equation}
where all functions depend on $u,\zeta,\xbar\zeta$, except where
explicitly indicated that the dependence on $u$ is replaced by a
dependence on $\widetilde u(u,\zeta,\xbar\zeta)$. 

In particular, this means that $\sigma^0_R(\widetilde
u),\dot{\xbar\sigma}^0_R(\widetilde u),\Psi^0_{aR}(\widetilde u),a=0,\dots,4$ and
$\Psi^0_{RiI}$, $i=0,1,2$, are invariant under complex
rescalings. Indeed, applying a complex rescaling to the non reduced
quantities amounts to applying the combined complex rescaling to the
reduced ones. In other words, only $P,\xbar P,\widetilde u$ change
while $\sigma^0_R,\dot{\xbar\sigma}^0_R,\Psi^0_{aR}$ are unchanged as
a function of their variables, while $\Psi^0_{iRI}$ are completely
unchanged. More generally, the transformation law of
$\sigma^0_R(\widetilde u),\dot{\xbar\sigma}^0_R(\widetilde
u),\Psi^0_{aR}(\widetilde
u),a=0,\dots,4$ under the extended BMS group combined
with complex rescalings simplifies to
\begin{equation}
  \label{eq:110}
\begin{split}
  &
  \sigma'^0_R=(\ddl{\zeta}{\zeta'})^{-\half}(\ddl{\xbar\zeta}{\xbar\zeta'})^{\frac{3}{2}}
  \Big[\sigma^0_R+\xbar\d^2\beta
  +\half\{\xbar\zeta',\xbar\zeta\}(\tilde u+\beta)\Big],\\
  & \dot{\xbar\sigma}'^0_R=(\ddl{\zeta}{\zeta'})^{2}[\dot{\xbar\sigma}^0_R
  +\half\{\zeta',\zeta\}\Big],\\
  &
  \Psi'^0_{4R}=(\ddl{\zeta}{\zeta'})^{\frac{5}{2}}(\ddl{\xbar\zeta}{\xbar\zeta'})^\half
  \Psi^0_{4R},\\
  &
  \Psi'^0_{3R}=(\ddl{\zeta}{\zeta'})^{2}\ddl{\xbar\zeta}{\xbar\zeta'}\Big[
  \Psi^0_{3R}-Y\Psi^0_{4R}\Big],\quad Y=\xbar\d\beta
  +\half\xbar\d\ln\ddl{\xbar\zeta'}{\xbar\zeta}(\tilde
  u+\beta),\\
&
  \Psi'^0_{2R}=(\ddl{\zeta}{\zeta'})^{\frac{3}{2}}
(\ddl{\xbar\zeta}{\xbar\zeta'})^{\frac{3}{2}}\Big[
  \Psi^0_{2R}-2Y\Psi^0_{3R}+Y^2\Psi^0_{4R}\Big],\\
&
  \Psi'^0_{1R}=\ddl{\zeta}{\zeta'}(\ddl{\xbar\zeta}{\xbar\zeta'})^{2}\Big[
  \Psi^0_{1R}-3Y\Psi^0_{2R}+3Y^2\Psi^0_{3R}-Y^3\Psi^0_{4R}\Big],\\
&
  \Psi'^0_{0R}=(\ddl{\zeta}{\zeta'})^\half(\ddl{\xbar\zeta}{\xbar\zeta'})^{\frac{5}{2}}
\Big[\Psi^0_{0R}-4Y\Psi^0_{1R}+6Y^2\Psi^0_{2R}-4Y^3\Psi^0_{3R}+Y^4\Psi^0_{4R}\Big], 
\end{split}
\end{equation}
where the primed quantities depend on $\widetilde
u'=J^{-\half}(\widetilde u+\beta),\zeta',\xbar\zeta'$, while the
unprimed ones depend on $\widetilde u,\zeta,\bar\zeta$. These
transformations simplify for the standard BMS group since the
Schwarzian derivative vanishes for this case. 

For the transformation law of $\widetilde\Psi^0_{iRI}$, we find
\begin{multline}
  \widetilde\Psi'^0_{2RI}=(\ddl{\zeta}{\zeta'})^{\frac{3}{2}}
(\ddl{\xbar\zeta}{\xbar\zeta'})^{\frac{3}{2}}\Big[\widetilde\Psi^0_{2RI}-\int^{0}_{-\beta}
  dv\,\big[\xbar\d\Psi^0_{3R}+\sigma^0_R\Psi^0_{4R}\big]
  +\\  +\big[-2\xbar\d\beta\Psi^0_{3R} +(\xbar\d\beta)^2\Psi^0_{4R}\big](-\beta)
\Big]. 
\end{multline}
\begin{multline}
  \widetilde\Psi'^0_{1RI}=\ddl{\zeta}{\zeta'}(\ddl{\xbar\zeta}{\xbar\zeta'})^{2}
  \Big[\widetilde\Psi^0_{1RI}-\int^{0}_{-\beta}
  dv\,\big[\xbar\d\Psi^0_{2R}+2\sigma^0_R\Psi^0_{3R}\big]
  +\\+\big[-3\xbar\d\beta\Psi^0_{2R}
  +3(\xbar\d\beta)^2\Psi^0_{3R}- (\xbar\d\beta)^3\Psi^0_{4R}\big](-\beta)
\Big], 
\end{multline}
\begin{multline}
  \widetilde\Psi'^0_{0RI}=(\ddl{\zeta}{\zeta'})^\half(\ddl{\xbar\zeta}{\xbar\zeta'})^{\frac{5}{2}}
\Big[\widetilde\Psi^0_{0RI}-\int^{0}_{-\beta}
  dv\,\big[\xbar\d\Psi^0_{1R}+3\sigma^0_R\Psi^0_{2R}\big]+\\
+\big[-4\xbar\d\beta\Psi^0_{1R}+6(\xbar\d\beta)^2\Psi^0_{2R}-4(\xbar\d\beta)^3\Psi^0_{3R}
+(\xbar\d\beta)^4\Psi^0_{4R}\big](-\beta)
\Big].
\end{multline}

The transformation laws of
$\Psi^0_{2RI},\Psi^0_{1RI}$ can be obtained from that of
$\widetilde\Psi^0_{2RI},\widetilde\Psi^0_{1RI}$ by using the first two
relations of \eqref{eq:116} and 
equation \eqref{eq:115a}, respectively \eqref{eq:115b} of
Appendix~\ref{sec:addit-transf-laws}. This gives
\begin{multline}
  \Psi'^0_{2RI}=(\ddl{\zeta}{\zeta'})^{\frac{3}{2}}
(\ddl{\xbar\zeta}{\xbar\zeta'})^{\frac{3}{2}}\Big[\Psi^0_{2RI}+\xbar\d^2\d^2\beta
+\half\{\xbar\zeta',\xbar\zeta\}(\xbar\sigma^0_R+\d^2\beta)
+\\+
\half\{\zeta',\zeta\}(\sigma^0_R+\xbar\d^2\beta)
-\int_{0}^{\tilde u} d\widetilde v\,
  \dot\sigma^0_R\dot{\xbar\sigma}^0_R\Big](-\beta), 
\end{multline}
\begin{multline}
  \Psi'^0_{1RI}=\ddl{\zeta}{\zeta'}(\ddl{\xbar\zeta}{\xbar\zeta'})^{2}\Big[
\Psi^0_{1RI}-\beta\xbar\d\Psi^0_{2RI}-3\xbar\d\beta\Psi^0_{2R}
+\frac{3}{2}\xbar\d^2\beta\xbar\d\xbar\sigma^0_R+\half\d^2\beta \xbar\d\sigma^0_R+\\+
\frac{3}{2}(\sigma^0_R+\xbar\d^2\beta)(\xbar\d\d^2\beta 
  -\xbar\d\beta\dot{\xbar\sigma}^0_R)+
  \half(\xbar\sigma^0_R+\d^2\beta)(\xbar\d^3\beta
  -\xbar\d\beta\dot{\sigma}^0_R)
-\\-3(\xbar\d\beta)^2\xbar\d\dot{\xbar\sigma}^0_R
+(\xbar\d\beta)^3\ddot{\xbar\sigma}^0_R+
\int_0^{\tilde u}d\widetilde v\,\xbar\d
\int^{\widetilde v}_0d\widetilde w\, \dot\sigma^0_R\dot{\xbar\sigma}^0_R+\\
+\int_0^{\tilde u}d\widetilde v \big[\half(\xbar\d\dot\sigma^0_R\xbar\sigma^0_R
-{3}\sigma^0_R\xbar\d \dot{\xbar\sigma}^0_R+{3}\dot\sigma^0_R\xbar\d \xbar\sigma^0_R
-\xbar\d\sigma^0_R \dot{\xbar\sigma}^0_R)-\xbar\d^3\xbar\sigma^0_R
\big] 
\Big](-\beta). 
\end{multline}

Finally, the transformation law of the Bondi mass aspect as chosen in
\eqref{eq:125} is given by \eqref{eq:121} after using
\eqref{eq:115c}. Note that the transformation law of the standard
expression for the Bondi mass aspect can easily be obtained by using
\eqref{eq:115}. 

\section{Discussion}
\label{sec:discussion}

In this work, we have generalized finite BMS$_4$ transformations to
include general holomorphic and antiholomorphic transformations as
well as time-dependent complex rescalings. A further interesting
generalization would be to abandon the reality conditions and consider
the transformations discussed in this work in the context of
$\cH$-space \cite{Newman1976,Ko:1977aa,Ko1981}.

The approach we have followed here is systematic and straightforward
but explicit computations are rather tedious and can presumably be
simplified in a more suitable set-up. Extracting physical consequences
from these transformation laws should be much more rewarding. We
conclude with some comments on why this should be the case.

The residual symmetry group we have investigated acts on the general
asymptotically flat solution space in the sense of Newman-Unti
\cite{Newman1962a}, containing not only the Kerr black hole
\cite{Kerr:1963ud} but also Robinson-Trautman waves
\cite{Robinson1960,Robinson1962}. In this context, the analog of the
time coordinate $\widetilde u$ used here has been introduced
previously in \cite{Adamo:2009vu} in order to express the latter
solutions in terms of a Bondi coordinate system where the conformal
factor is the one for the unit sphere.

The transformations also naturally act on the would-be conserved BMS
currents including Bondi mass and angular momentum aspects, which are
built out of the data considered here. In order to cover the most
general case, the expressions considered for instance in
\cite{Barnich:2013axa} have first to be generalized to the case of a
variable, complex, $u$-dependent factor $P$. This will be done in
\cite{Barnich:2016}. 

The relevance of the transformation formulas to the gravitational
memory effect \cite{Zeldovich1974,Christodoulou:1991cr} as described
in \cite{Frauendiener1992} (see
\cite{Strominger:2014pwa,Pasterski:2015tva} for recent discussions) is
obvious. The question of what part of this effect is controlled by BMS
transformations boils down to a question about suitable orbits of the
BMS group. These problems will be discussed in more details elsewhere,
together with other applications involving topology-changing mappings.

\section*{Acknowledgements}
\label{sec:acknowledgements}

\addcontentsline{toc}{section}{Acknowledgments}

This work is supported in part by the Fund for Scientific
Research-FNRS (Belgium), by IISN-Belgium, and by ``Communaut\'e fran\c
caise de Belgique - Actions de Recherche Concert\'ees''. C.~Troessaert
is Conicyt (Fondecyt postdoctoral grant 3140125) and Laurent Houart
postdoctoral fellow. The Centro de Estudios Cient\'ificos (CECs) is
funded by the Chilean Government through the Centers of Excellence
Base Financing Program of Conicyt. The authors thank Pujian Mao for
pointing out relevant references on the memory effect. They are
grateful to Per Sundell and his collaborators at the Universidad
Andr\'es Bello (Chile) for hospitality during the final stages of this
work.

\appendix

\section{Newman-Penrose field equations in 3d}
\label{sec:appEOM3D}

The Einstein equations in three dimensions can be expressed as 
\begin{eqnarray}
\label{eq:NPeq3DII}
D\sigma - \delta \kappa & = &  (\epsilon+2\sigma)\sigma - (\tau-\pi+2\beta) \kappa ,\\
	\label{eq:NPeq3DI}
	D\tau - \Delta \kappa & = &  2 (\tau+ \pi)\sigma- 2 \kappa\gamma,\\
	\label{eq:NPeq3DIV}
	D\beta - \delta \epsilon & = &  2 (\beta+\pi) \sigma  - (2 \mu +
        \gamma)\kappa - (\beta-\pi) \epsilon ,\\
\label{eq:NPeq3DIII}
	D\gamma - \Delta \epsilon  & = & 2(\tau+\pi) \beta- 2
        \epsilon\gamma + 2 \tau \pi - 2 \kappa
	\nu  +\frac{1}{L^2},\\
\label{eq:NPeq3DVI}
	D\mu - \delta \pi & = & (2 \sigma - \epsilon) \mu +  \pi^2 
	-\kappa \nu -\frac{1}{2 L^2} ,\\
	\label{eq:NPeq3DV}
	D\nu - \Delta \pi & = &   2 (\pi+\tau)\mu - 2 \epsilon \nu,\\
	\label{eq:NPeq3DIX}
	 \Delta\mu - \delta \nu & = &   - (2 \mu +\gamma) \mu
	+  (\pi +2 \beta- \tau)\nu,\\
	\label{eq:NPeq3DVIII}
	\Delta\beta - \delta \gamma & = & (\beta-\tau)\gamma - 2 (\beta+\tau)\mu
	  + (2\sigma+\epsilon) \nu,\\
	\label{eq:NPeq3DVII}
	\Delta\sigma - \delta \tau & = &  (\gamma- 2 \mu)\sigma - \tau^2
	+  \nu \kappa +\frac{1}{2 L^2},
\end{eqnarray}
while vanishing of torsion can be written as 
\begin{eqnarray}
	\label{eq:NPeq3DX}
	D n^\mu - \Delta l^\mu & = & -\gamma l^\mu - \epsilon n^\mu + 2 (\pi + \tau)
	m^\mu,\\
	\label{eq:NPeq3DXI}
	D m^\mu - \delta l^\mu & = & (\pi - \beta) l^\mu - \kappa n^\mu +2\sigma m^\mu,\\
	\label{eq:NPeq3DXII}
	\Delta m^\mu - \delta n^\mu & = & \nu l^\mu + (\beta - \tau ) n^\mu -2 \mu m^\mu.
\end{eqnarray}

\section{Additional transformation laws in 4d}
\label{sec:addit-transf-laws}

\begin{multline}
  \xbar\d'^2 \xbar{\sigma}'^0_R=J^{\frac{3}{2}}\Big[[\xbar\d^2
+\half\{\xbar\zeta',\xbar\zeta\}]\xbar\sigma^0_R
+[\xbar\d^2+\half\{\xbar\zeta',\xbar\zeta\}]\d^2\beta\\
-[\xbar\d^2\beta
+\half\{\xbar\zeta',\xbar\zeta\}(\widetilde u+\beta)]
\dot{\xbar\sigma}^0_R
-2Y\xbar\d\dot{\xbar\sigma}^0_R
+Y^2\ddot{\xbar\sigma}^0_R\Big], \label{eq:115} 
\end{multline}
\begin{multline}
  \label{eq:115a}
  \xbar\d'^2\xbar\sigma'^0_R+\sigma'^0_R\dot{\xbar\sigma}'^0_R=
J^{\frac{3}{2}}\Big[\xbar\d^2\xbar\sigma^0_R+\sigma^0_R\dot{\xbar\sigma}^0_R
-2Y\xbar\d\dot{\xbar\sigma}^0_R
+Y^2\ddot{\xbar\sigma}^0_R+\half\{\xbar\zeta',\xbar\zeta\}({\xbar\sigma}^0_R+\d^2\beta)
+\\+\half\{\zeta',\zeta\}({\sigma}^0_R+\xbar\d^2\beta)+\d^2\xbar\d^2\beta
+\frac{1}{4}\{\xbar\zeta',\xbar\zeta\}\{\zeta',\zeta\}(\widetilde u+\beta)\Big],
\end{multline}
\begin{multline}
  \sigma'^0_R\xbar\d'\xbar\sigma'^0_R+\half\xbar\d'(\sigma'^0_R\xbar\sigma'^0_R)=
  \ddl{\zeta}{\zeta'} (\ddl{\xbar\zeta}{\xbar\zeta'})^{2}\Big[
  \frac{3}{2}\big[\sigma^0_R +\xbar\d^2\beta+\half
  \{\xbar\zeta',\xbar\zeta\}(\widetilde
  u+\beta)\big]\\
\big[\xbar\d\xbar\sigma^0_R+\xbar\d\d^2\beta
  -Y\dot{\xbar\sigma}^0_R-\frac{1}{4}\xbar\d\ln\ddl{\xbar\zeta'}{\xbar\zeta}
  \{\zeta',\zeta\}(\widetilde u+\beta)\big]+\\+
  \half\big[\xbar\sigma^0_R+\d^2\beta+
  \half  \{\zeta',\zeta\}(\widetilde u+\beta)\big]\\
\big[\xbar\d\sigma^0_R+\xbar\d^3\beta
  -Y\dot{\sigma}^0_R+\frac{1}{2}(\xbar\d-\half \xbar\d\ln\ddl{\xbar\zeta'}{\xbar\zeta})
  \{\xbar\zeta',\xbar\zeta\}(\widetilde u+\beta)\big] \Big], \label{eq:115b}
\end{multline}
\begin{multline}
  \label{eq:115c}
  (\int^{\widetilde u}_0d\widetilde v\,
  \dot\sigma^0_R\dot{\xbar\sigma}^0_R)'=J^{\frac{3}{2}}
\Big[\half\{\xbar\zeta',\xbar\zeta\}(\xbar\sigma^0_R-\xbar\sigma^0_R(-\beta))
+ \half\{\zeta',\zeta\}(\sigma^0_R
-\sigma^0_R(-\beta))+\\+\frac{1}{4}\{\xbar\zeta',\xbar\zeta\}
\{\zeta',\zeta\}(\widetilde u+\beta)
  +\int^{\widetilde u}_{-\beta} d\widetilde v\,
  \dot\sigma^0_R\dot{\xbar\sigma}^0_R\Big].
\end{multline}

\addcontentsline{toc}{section}{References}

%\bibliography{/Users/gbarnich/Dropbox/Literature/master}
%\bibliography{C:/Users/Glenn/Dropbox/Literature/master}

\def\cprime{$'$}
\providecommand{\href}[2]{#2}\begingroup\raggedright\endgroup

\end{document}